\documentclass[journal]{IEEEtran}
\usepackage[utf8]{inputenc}
\usepackage{textcomp}
\usepackage{eurosym}
\usepackage{verbatim}
\usepackage{cite}
\usepackage{amsmath,amssymb,amsfonts}
\usepackage{algorithmic}
\usepackage{graphicx}
\usepackage[super]{nth}
\usepackage{textcomp}
\usepackage{xcolor}
\usepackage{xfrac}
\usepackage{multicol}
\usepackage{multirow}
\usepackage{booktabs}
\usepackage{tabularx, ragged2e}
\usepackage{enumitem} 
\usepackage{gensymb}
\usepackage{bm}
\usepackage{lipsum}
\usepackage{esvect}
\usepackage{siunitx}
\usepackage{tikz}
\usepackage{forest,array}
\usetikzlibrary{arrows.meta, trees, positioning, shapes, arrows, fadings, mindmap, backgrounds, decorations.text, patterns, 3d, angles, calc, intersections, quotes}
\usepackage{placeins}
\usepackage{afterpage}
\usepackage{pgfplots}
\pgfplotsset{compat=1.14}

\newcolumntype{C}{>{\Centering\arraybackslash}X}

\tikzset{pics/wedge/.style={code={%
\tikzset{wedge/.cd,#1}
\def\kvw##1{\pgfkeysvalueof{/tikz/wedge/##1}}
\pgfmathtruncatemacro{\itest}{3*(1+sign(sin(\kvw{alpha})))+1+sign(sin(\kvw{beta}))}
    \ifcase\itest
       \draw[fill=\kvw{color},very thin]
                       (\kvw{alpha}:\kvw{radius}) 
                       -- ++(0,-\kvw{h}) arc(\kvw{alpha}:\kvw{beta}:\kvw{radius}) 
                       -- ++(0,\kvw{h})  arc(\kvw{beta}:\kvw{alpha}:\kvw{radius});
   \or
       \draw[fill=\kvw{color},very thin]
                       (\kvw{alpha}:\kvw{radius}) 
                       -- ++(0,-\kvw{h}) arc(\kvw{alpha}:\kvw{beta}:\kvw{radius}) 
                       -- ++(0,\kvw{h})  arc(\kvw{beta}:\kvw{alpha}:\kvw{radius});
   \or
       \draw[fill=\kvw{color},very thin]
                       (\kvw{alpha}:\kvw{radius}) 
                       -- ++(0,-\kvw{h}) arc(\kvw{alpha}:360:\kvw{radius}) 
                       -- ++(0,\kvw{h})  arc(360:\kvw{alpha}:\kvw{radius});
   \or
       \draw[fill=\kvw{color},very thin]
                       (180:\kvw{radius}) 
                       -- ++(0,-\kvw{h}) arc(180:\kvw{beta}:\kvw{radius}) 
                       -- ++(0,\kvw{h})  arc(\kvw{beta}:180:\kvw{radius});
   \or
       \draw[fill=\kvw{color},very thin]
                       (180:\kvw{radius}) 
                       -- ++(0,-\kvw{h}) arc(180:0:\kvw{radius}) 
                       -- ++(0,\kvw{h})  arc(0:180:\kvw{radius});
   \or
   \or
       \draw[fill=\kvw{color},very thin]
                       (180:\kvw{radius}) 
                       -- ++(0,-\kvw{h}) arc(180:\kvw{beta}:\kvw{radius}) 
                       -- ++(0,\kvw{h})  arc(\kvw{beta}:180:\kvw{radius});
   \or
    \pgfmathtruncatemacro{\ibeta}{sign(cos(\kvw{beta}))}
    \ifnum\ibeta=1
        \draw[fill=\kvw{color},very thin] 
                        (180:\kvw{radius}) 
                        -- ++(0,-\kvw{h}) arc(180:360:\kvw{radius}) 
                        -- ++(0,\kvw{h})  arc(360:180:\kvw{radius});
    \fi
    \or
    \pgfmathtruncatemacro{\ibeta}{sign(sin(\kvw{alpha})-sin(\kvw{beta}))}
    \ifnum\ibeta=1
       \draw[fill=\kvw{color},very thin]
                        (180:\kvw{radius}) 
                        -- ++(0,-\kvw{h}) arc(180:360:\kvw{radius}) 
                        -- ++(0,\kvw{h})  arc(360:180:\kvw{radius});
    \fi
    \fi
    \path[fill=\kvw{color},draw=black] (0,0)--
    (\kvw{alpha}:\kvw{radius})  arc(\kvw{alpha}:\kvw{beta}:\kvw{radius})
                                     --cycle;
    }},
wedge/.cd,alpha/.initial=0,beta/.initial=0,
color/.initial=blue,
mix color/.initial=gray,radius/.initial=1cm,h/.initial=1cm,
/tikz/.cd,
pics/3d pie chart/.style={code={
    \def\kvw##1{\pgfkeysvalueof{/tikz/3d pie chart/##1}}
    \begin{scope}[yscale=\kvw{aspect},transform shape]
    \path[preaction={fill=black,opacity=.8,
        path fading=circle with fuzzy edge 10 percent}] 
        (0,-\kvw{h}-\kvw{radius}/6.5) 
        circle[radius=1.05*\kvw{radius}];
    \pgfmathsetmacro{\mysum}{0}      
    \foreach \XX/\ZZ  in {#1}  
    {\pgfmathsetmacro{\mysum}{\mysum+\XX}
    \xdef\mysum{\mysum}}
    \pgfmathsetmacro{\myangle}{\kvw{alpha0}}
    \foreach \XX/\ZZ [count=\YY starting from 0,remember=\myangle as \myangle] in {#1} 
    {\pgfmathsetmacro{\myangleB}{\myangle+\XX*(360/\mysum)}
    \pgfmathsetmacro{\mycolor}{{\kvw{colors}}[\YY]}
    \pic{wedge={alpha=\myangle,beta=\myangleB,color=\mycolor,
        radius/.expanded=\kvw{radius},
        h/.expanded=\kvw{h}
        }};
    \fill (\myangle/2+\myangleB/2:\kvw{radius}*\kvw{eccentricity})
    coordinate (\kvw{cname}-\YY) circle[radius=3pt];
    \pgfmathtruncatemacro{\mysign}{sign(cos(\myangle/2+\myangleB/2))} 
    \draw[thick] (\kvw{cname}-\YY)  -- 
    ++(\myangle/2+\myangleB/2:\kvw{armA}) -- ++ 
    (\mysign*1.4,0)
    \ifnum\mysign<0
        node[above right,transform shape=false]{\ZZ}
        node[below right,transform shape=false]{\XX\%}
    \else
        node[above left,transform shape=false]{\ZZ}
        node[below left,transform shape=false]{\XX\%}
    \fi;    
    \pgfmathsetmacro{\myangle}{\myangleB}
    }
    \end{scope}                       
    }},
    3d pie chart/.cd,
    colors/.initial={"blue","red","orange","green","yellow"},
    radius/.initial=0.1cm,h/.initial=1cm,alpha0/.initial=0,
    aspect/.initial=0.5,eccentricity/.initial=0.7,cname/.initial=c,
    armA/.initial=1.3cm
}

\def\BibTeX{{\rm B\kern-.05em{\sc i\kern-.025em b}\kern-.08em
    T\kern-.1667em\lower.7ex\hbox{E}\kern-.125emX}}
    
\ifCLASSINFOpdf
\else
\fi

\begin{document}

\title{Vehicle Telematics Via Exteroceptive Sensors:\\ A Survey}

\author{Fernando~Molano~Ortiz, 
        Matteo~Sammarco, 
        Luís~Henrique~M.~K.~Costa,~\IEEEmembership{Senior Member,~IEEE}
        and~Marcin~Detyniecki
\thanks{F. M. Ortiz and L. H. M. K. Costa are with the GTA/PEE-COPPE/DEL-Poli, Federal University of Rio de Janeiro, Brazil, (e-mail: \{fmolano, luish\}@gta.ufrj.br)}
\thanks{M. Sammarco and M. Detyniecki are with Axa, France (e-mail: {\textit{\{matteo.sammarco, marcin.detyniecki\}}@axa.com}).}}

\markboth{}%
{Shell \MakeLowercase{\textit{et al.}}: Bare Demo of IEEEtran.cls for IEEE Journals}

\maketitle

\begin{abstract}
Whereas a very large number of sensors are available in the automotive field, currently just a few of them, mostly proprioceptive ones, are used in telematics, automotive insurance, and mobility safety research.
In this paper, we show that exteroceptive sensors, like microphones or cameras, could replace proprioceptive ones in many fields. 
Our main motivation is to provide the reader with alternative ideas for the development of telematics applications when proprioceptive sensors are unusable for technological issues, privacy concerns, or lack of availability in commercial devices.
We first introduce a taxonomy of sensors in telematics. 
Then, we review in detail all exteroceptive sensors of some interest for vehicle telematics, highlighting advantages, drawbacks, and availability in off-the-shelf devices.
Successively, we present a list of notable telematics services and applications in research and industry like driving profiling or vehicular safety. 
For each of them, we report the most recent and important works relying on exteroceptive sensors, as long as the available datasets.
We conclude showing open challenges using exteroceptive sensors both for industry and research.
\end{abstract}


\section{Introduction}
\label{sec:introduction}

\IEEEPARstart{V}{ehicle} telematics is steadily evolving as a solution to improve mobility safety, vehicles efficiency, and maintenance~\cite{Markets:2020}. Sensors which monitor mechanical, electrical, and electronic systems of the vehicle produce information collected by an Electronic Control Units (ECUs) which optimizes vehicle performance and enhances safety by producing preventive maintenance reports. Naturally, correct functioning of the machine alone is not enough to prevent accidents. The World Health Organization reports that road traffic accidents represent one of the leading causes of global deaths~\cite{WHO:2018}, while the European Road Safety Observatory quantifies the socioeconomic consequences for traffic injuries in 2018 as \euro\,120 billion~\cite{EU:2019}.

As a logical consequence, industry and researchers pursue innovative methods to support drivers, through Advanced Driver Assistance Systems (ADAS) based on external and internal vehicle sensing. External sensing allows gathering information on the environment around the vehicle through specific sensors, with the possibility of sharing this information with other vehicles in proximity~\cite{Alsultan:2014}. As a result, vehicle telematics has become more relevant, as it directly impacts drivers, passengers and the environment around the vehicle. 
Figure\,\ref{fig:automotive_market} shows a projection of the electronics market growth between 2018 and 2023~\cite{Gartner_Forecast:2018}. Automotive electronics forecast is only comparable to industrial electronics, pushed by the fourth industrial revolution.

\afterpage{
\begin{figure}[t]
\centering
\hspace{-0.5cm}
\resizebox{0.42\textwidth}{!}{

\begin{tikzpicture}
\begin{axis}[
    xbar stacked, 
    xmin=0, xmax=10,
    axis x line*=bottom,
    axis y line*=none,
    xlabel={Percentage \%},
    symbolic y coords={%
        {Industrial},
        {Automotive},
        {Military/Civil Aerospace},
        {Total IC},
        {Consumer},
        {Communication},
        {Data Processing}},
    y tick label style={text width=2.5cm,align=right},
    ytick=data,
    ytick=data,
    bar width=6mm,
    ]
    \addplot[left color=black!20!blue, right color=black!40!red]coordinates {
        (8.60,{Industrial})
        (8.80,{Automotive}) 
        (2.15,{Military/Civil Aerospace}) 
        (3.10,{Total IC}) 
        (2.20,{Consumer})
        (1.50,{Communication})
        (0.98,{Data Processing})};
\end{axis}   
\begin{scope}[shift={(5.4,4.3)}]
    \path[3d pie chart/.cd,radius=1.5cm,h=0.5cm,colors={"blue!25","black!10!yellow","purple!60","red!90","black!30!green!50"}] 
    pic{3d pie chart={18/Others,36/LiDAR,22/Imaging,15/Radar,9/Ultrasonic}};
\end{scope}
\begin{scope}
    \filldraw[black] (5.8,1.25) circle (0.1cm) node[] (auto) {};
    \draw (5.8,1.255) -- (5.8,1.8) node[line width=0.5mm] {}; 
    \node at (5.8,2.5) {CAGR for Automotive Sensing};
    \node at (5.8,2.1) {Market (2016-2022)~\cite{Yole_Sensor_Forecast:2017}};
    \node [draw, thick, shape=rectangle, minimum width=7cm, minimum height=4cm, anchor=center] at (5.8,3.8) {};
\end{scope}
\end{tikzpicture}
}
    \caption{Electronics market growth projection (2018-2023 CAGR)~\cite{Gartner_Forecast:2018}. The pie chart shows the prediction of automotive sensors market growth (2016-2022 CAGR)~\cite{Yole_Sensor_Forecast:2017}, focused on exteroceptive sensors.}
	\label{fig:automotive_market}
\end{figure}
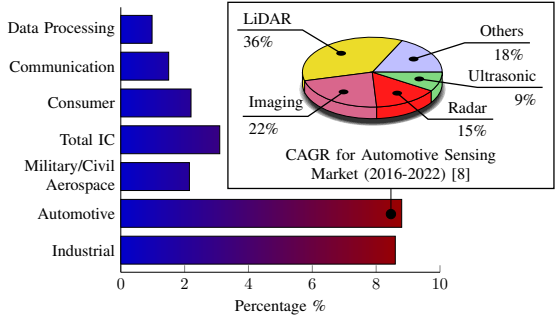
}

Vehicles are today equipped with a myriad of sensors, which integrate different systems and help to improve, adapt, or automate vehicle safety and driving experience. These systems assist drivers by offering precautions to reduce risk exposure or by cooperatively automating driving tasks, with the aim of minimizing human errors~\cite{Guo:2019}. Typically, counting only on measurements from internal (``proprioceptive'') sensors is not sufficient to provide safety and warning applications associated with the external environment. With exteroceptive sensors instead, vehicles have the ability to acquire information on the surrounding environment, recognizing other factors and objects that coexist in the same space. Vehicle external sensing is gaining importance especially with the proliferation of cameras which, combined with the improved image processing and analysis, enables a wide range of applications~\cite{IHS:2018}. Consequently, imaging is among the areas with the highest projection in automotive electronics as shown in Figure~\ref{fig:automotive_market}. LiDARs have the primacy as exteroceptive sensors expected to be the most demanded in automotive. Unlike cameras, they provide an omnidirectional sensing and they do not suffer from scarce light conditions. Radar and ultrasonic sensors share another quarter of the market, while 18\% of the market will be taken by other exteroceptive sensors like microphones.

Data generated by exteroceptive sensors is of primary interest for drivers and passenger. In fact, the environmental sensing is fundamental for ADAS, collision avoidance and safety applications. Nonetheless, the same data has a valence for smart cities, e.g., sensing the road conditions, and for insurance companies to establish driving profiles and calculate insurance premiums.

\smallskip\textit{Contribution and Outline.}
In this paper we present an overview of exteroceptive sensors and their use in vehicle telematics along with relative services and applications. Our main focus is on the safety application area, but we also cover other important fields related to mobility, like navigation, road monitoring and driving behavior analysis. Such applications are interesting for car manufacturers, insurance companies, smart cities, as well as drivers and passengers. The purpose it to provide a clear taxonomy of works per telematics application and per exteroceptive sensor. Studies are mainly selected proportional to their relevance, novelty, and publication date. In this process, we provide background information on sensors in telematics, detailing and comparing the exteroceptive sensors particularly. When applicable, we describe Original Equipment Manufacturer (OEM) and aftermarket devices which include such sensors.

\begin{figure}
\hspace{-0.5cm}
\centering
\resizebox{0.5\textwidth}{!}{
\tikzstyle{section}=[align=center, text width=1.9cm, minimum height=0.8cm, rectangle, draw = black, rounded corners = 3mm]
\tikzstyle{subsec}=[align=left, text width=3.75cm, minimum height=0.5cm, rectangle, draw = black, rounded corners = 3mm]
\tikzstyle{subsubsec}=[align=left, text width=4.55cm, minimum height=0.5cm, rectangle, draw = black, rounded corners = 3mm]
\begin{tikzpicture}[thick,scale=1, every node/.style={scale=1.1}]

\draw (0,10) node[section] (sec_1) {\ref{sec:introduction} Introduction};
\draw (0,8.2) node[section] (sec_2) {\ref{sec:background} Background};
\draw (0,5.05) node[section] (sec_3) {\ref{sec:_ext_sensors_telematics}\\Exteroceptive Sensors for Telematics};
\draw (0,1.45) node[section] (sec_4) {\ref{sec:services_and_applications}\\Services and Applications};
\draw (0,-1.3) node[section] (sec_5) {\ref{sec:challenges}\\Open Research/ Challenges};
\draw (0,-3.15) node[section] (sec_6) {\ref{sec:conclusion} Conclusion};

\draw (3.7,10) node[subsec] (subsec_intro) {$-$ Overview \\$-$ Contribution};
\draw (3.7,8.8) node[subsec] (subsec_classification) {\ref{subsec:classification} Proprioceptive \textit{vs.} \hspace*{0.6cm} exteroceptive sensors};
\draw (3.7,7.6) node[subsec] (subsec_OTS) {\ref{subsec:OTS_devices} OTS telematics \hspace*{0.6cm} devices};

\draw (3.7,5.05) node[subsec] (subsec_exteroceptive) {\ref{subsec:gnss} GNSS\\ 
    \ref{subsec:magnetometer} Magnetometer\\ 
    \ref{subsec:microphone} Microphone\\ 
    \ref{subsec:biometric} Biometric sensors\\ 
    \ref{subsec:ultrasonic} Ultrasonic sensor\\ 
    \ref{subsec:radar} Radar\\ 
    \ref{subsec:lidar} LiDAR\\ 
    \ref{subsec:camera} Camera};

\draw (3.7,2.7) node[subsec] (subsec_safety) {\ref{subsec:safety} Safety};
\draw (3.7,1.95) node[subsec] (subsec_driv) {\ref{subsec:driving_behavior} Driving behavior};
\draw (3.7,1.15) node[subsec] (subsec_road) {\ref{subsec:road_monitoring} Road monitoring};
\draw (3.7,0.35) node[subsec] (subsec_nav) {\ref{subsec:navigation} Navigation};
\draw (3.7,-1.3) node[subsec] (subsec_challenges) {$-$ Which data is important?\\ 
    $-$ Data processing\\ 
    $-$ Security\\ 
    $-$ Risk assessment};
\draw (3.7,-3.15) node[subsec] (subsec_conclusion) {$-$ Summary \\$-$ Perspectives};


\draw (9,10.2) node[subsubsec] (subsubsec_pass_act) {\ref{subsubsec:classification_active_passive} Active/passive sensors};

\draw (9,7.85) node[subsubsec] (subsubsec_OTS) {\ref{subsubsec:OBD_CAN} OBD-II dongle and \hspace*{0.8cm} CAN bus readers\\ 
    \ref{subsubsec:digital_tachograph} Digital tachograph\\ 
    \ref{subssubsec:bbox_wind} Black-box and \hspace*{0.8cm} windshield devices\\ 
    \ref{subsubsec:dashcam} Dashcam\\ 
    \ref{subsubsec:smartphones} Smartphones\\ 
    \ref{subsubsec:wearable} Wearable devices};
\draw (9,4.8) node[subsubsec] (subsubsec_safety) {\ref{subsubsec:tire_wear} Tire wear\\ 
    \ref{subsubsec:crash_detection} Collision detection\\ 
    \ref{subsubsec:collision_avoidance} Collision avoidance\\ 
    \ref{subsubsec:Lane_departure} Lane departure warning};
\draw (9,2.4) node[subsubsec] (subsubsec_driv) {\ref{subsubsec:driving_profiling} Driving profiling\\ 
    \ref{subsubsec:driver_passenger_identification} Driver detection\\ 
    \ref{subsubsec:driver_identification} Driver identification\\ 
    \ref{subsubsec:Health_monitoring} Driver health monitoring\\ 
    \ref{subsubsec:Driving_distraction} Driving distractions};
\draw (9,-0.25) node[subsubsec] (subsubsec_road) {\ref{subsubsec:road_porosity} Road porosity\\ 
    \ref{subsubsec:road_wetness} Road wetness\\ 
    \ref{subsubsec:pothole_detection} Pothole detection\\ 
    \ref{subsubsec:road_type_classification} Road type classification\\ 
    \ref{subsubsec:parking_space_detection} Parking space detection};
\draw (9,-2.7) node[subsubsec] (subsubsec_nav) {\ref{subsubsec:GNSS-based} GNSS/INS-based \hspace*{0.85cm} navigation \\ 
    \ref{subsubsec:SLAM-based} SLAM-based navigation\\ 
    \ref{subsubsec:map_tracking} Map tracking datasets};

\foreach \f/\t in
    {sec_1/subsec_intro, sec_2/subsec_OTS, sec_2/subsec_classification, sec_3/subsec_exteroceptive, sec_4/subsec_nav, sec_4/subsec_road, sec_4/subsec_driv, sec_4/subsec_safety, sec_5/subsec_challenges, sec_6/subsec_conclusion, subsec_classification/subsubsec_pass_act, subsec_OTS/subsubsec_OTS, subsec_safety/subsubsec_safety, subsec_driv/subsubsec_driv, subsec_road/subsubsec_road, subsec_nav/subsubsec_nav}
    \draw[black, very thick] (\f.east) -- (\t.west);
    \end{tikzpicture}}
	\caption{Paper organization outline.}
	\label{fig:outline}
\end{figure}
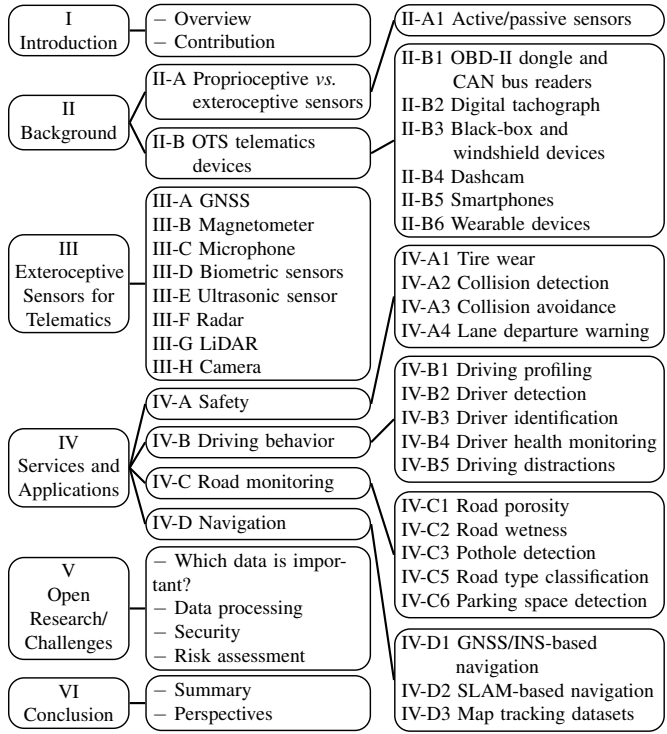

For the sake of readability, the organization of this paper is shown in Figure\,\ref{fig:outline}. 
Section\,\ref{sec:background} reviews sensor aspects in telematics, their classification, advantages and disadvantages. It also provides an overview of Off-the-Shelf (OTS) telematics devices. 
Section\,\ref{sec:_ext_sensors_telematics} details exteroceptive sensors used in vehicle telematics. Section\,\ref{sec:services_and_applications} shows a taxonomy of high level telematics applications like navigation, road monitoring, driving behavior, and safety. For each application, a list of works by sensor is provided.
Open research and challenges in the usage of exteroceptive sensors for vehicular telematics are presented in Section\,\ref{sec:challenges}, while Section\,\ref{sec:conclusion} concludes the survey with summary and perspectives.

\section{Background}
\label{sec:background}

\begin{figure}[!t]
    \centering
    \resizebox{0.452\textwidth}{!}{
    \begin{tikzpicture}
    
    \tikzstyle{Perception}=[align=center, text width=2cm, minimum height=1.5cm, rectangle, draw = black, rounded corners = 2mm]
    \tikzstyle{Sensor}=[align=center, text width=2.3cm, minimum height=0.5cm, rectangle, draw = black, rounded corners = 1mm]
    \tikzstyle{Standalone}=[align=center, text width=1.7cm, minimum height=1cm, rectangle, draw = black, rounded corners = 2mm]
    \tikzstyle{Services}=[align=center, text width=2.3cm, minimum height=1cm, rectangle, draw = black, rounded corners = 2mm]

    \draw (0,-.1) node[align=center, text width=1.95cm, minimum height=7.3cm, rectangle, draw = black, rounded corners = 3mm, dashed] (g_sensor) {};
    \draw (0,4.0) node[align=center, text width=2.5cm] (t_g_sensor) {\textit{OTS Telematics devices}};
    \draw (0,2.9) node[Standalone, fill=orange!40] (bbox) {Black-box / Windshield};
    \draw (0,1.7) node[Standalone, fill=orange!40] (dash) {Dashcam};
    \draw (0,0.5) node[Standalone, fill=orange!40] (smart) {Smartphone};
    \draw (0,-0.7) node[Standalone, fill=orange!40] (wear) {Wearable};
    \draw (0,-1.9) node[Standalone, fill=orange!40] (obd) {OBD-II dongle};
    \draw (0,-3.1) node[Standalone, fill=orange!40] (tach) {Tachograph};
    \draw (4,1.68) node[fill=cyan!30, opacity=0.8, align=center, text width=2.5cm, minimum height=5.35cm, rectangle, draw = black, rounded corners = 2mm, dashed] (g_sensor) {};
    \draw (4,4.55) node[align=center] (t_g_sensor) {\textit{Exteroceptive}};
    \draw (4,4) node[Sensor, fill=gray!20] (ult) {Ultrasonic};
    \draw (4,3.35) node[Sensor, fill=gray!20] (rad) {Radar};
    \draw (4,2.7) node[Sensor, fill=gray!20] (lid) {LiDAR};
    \draw (4,2.05) node[Sensor, fill=white] (cam) {Camera};
    \draw (4,1.4) node[Sensor, fill=white] (mic) {Microphone};
    \draw (4,0.725) node[Sensor, fill=gray!20] (gnss) {GNSS};
    \draw (4,0.05) node[Sensor, fill=white] (mag) {Magnetometer};
    \draw (4,-0.62) node[Sensor, fill=white] (plet) {Biometric};
    
    \draw (4,-3.15) node[fill=yellow!30, opacity=0.8, align=center, text width=2.5cm, minimum height=3.1cm, rectangle, draw = black, rounded corners = 2mm, dashed] (g_sensor) {};
    \draw (4,-1.4) node[align=center] (t_g_sensor) {\textit{Proprioceptive}};
    \draw (4,-2) node[Sensor, fill=white] (gyr) {Gyroscope};
    \draw (4,-2.67) node[Sensor, fill=white] (acc) {Accelerometer};
    \draw (4,-3.5) node[Sensor, fill=white] (can) {CAN bus sensors}; 
    \draw (4,-4.35) node[Sensor, fill=white] (hall) {Hall effect};
    
    \draw (8,0.6) node[align=center, text width=2.5cm, minimum height=6.15cm, rectangle, draw = black, rounded corners = 3mm, dashed] (g_sensor) {};
    \draw (8,4.1) node[align=center, text width=3cm] (t_g_sensor) {\textit{Services and applications}};
    \draw (8,3) node[Services, fill=green!40] (safe) {Safety (Sec.~\ref{subsec:safety})};
    \draw (8,1.45) node[Services, fill=green!40] (driv) {Driving behavior (Sec.~\ref{subsec:driving_behavior})};
    \draw (8,-0.25) node[Services, fill=green!40] (road) {Road monitoring (Sec.~\ref{subsec:road_monitoring})};
    \draw (8,-1.8) node[Services, fill=green!40] (nav) {Navigation (Sec.~\ref{subsec:navigation})};
    
    \foreach \f/\t in
        {dash/cam, smart/cam/, smart/mic, bbox/gnss, smart/gnss, wear/plet, smart/mag, wear/gyr, smart/gyr, bbox/acc, wear/acc, smart/acc, tach/hall, ult/road, ult/safe, rad/road, rad/nav, rad/safe, lid/nav, lid/safe, cam/nav, cam/road, cam/driv, cam/safe, mic/driv, mic/road, gnss/nav, gnss/driv, mag/nav, plet/driv, obd/can, obd/acc, obd/gnss, dash/mic, dash/acc, obd/mag}
        \draw[black, very thick] (\f.east) -- (\t.west);
    \end{tikzpicture}  
    }
\caption{{Bipartite graphs showing the relationship between the most widely used sensors in vehicular telematics  (in the middle), OTS telematics devices (on the left), and between sensors and telematics services and applications (on the right). 
In the middle, exteroceptive sensors are nodes on the top, while proprioceptive on the bottom.
Active sensors are represented as gray colored nodes, passive sensors as white colored nodes.}}
\label{fig:built-in_sensors}
\end{figure}
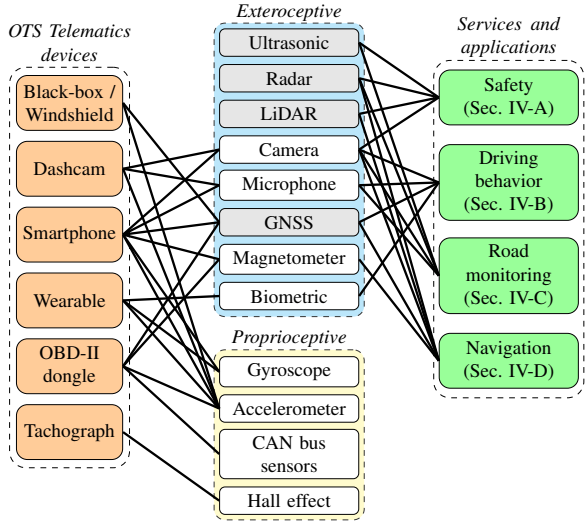

Sensors in telematics enable monitoring a broad range of functions inherent to the management of diverse driving activities. Electronic sensing systems and data processing capacity reduce driver's workload and provide innovative services. This section presents a classification of sensors for telematics purposes, according to the environment in which they operate. Figure~\ref{fig:built-in_sensors} illustrates the proposed classification where sensors are placed in the central column. On the left-hand side, each sensor is connected to OTS telematics devices where it is embedded, while at the right-hand side possible fields of application are identified.

\subsection{Proprioceptive vs. exteroceptive sensors}
\label{subsec:classification}

A wide variety of sensors is used in regular vehicles, the majority of them to gather information on internal mechanisms. Self-driving vehicles on the other hand incorporate external sensors whose function is critical to analyze the surrounding environment. Therefore, vehicular telematics is no longer merely mechanical, leading to the analysis of internal and external variables. As such, a basic classification of the sensors is according to the sensed variables, as \textit{proprioceptive} or \textit{exteroceptive}~\cite{Siegwart:2004}.

\textit{Proprioceptive sensors} measure variations in signals generated by the vehicle's internal systems (motor speed, battery level, etc.). Those measurements allow estimating different metrics specific to the vehicle, such as speed, fluid levels, acceleration, among other topics of interest for vehicle telematics. An accelerometer is an example of proprioceptive sensor.

\textit{Exteroceptive sensors} allow vehicles to be in contact with stimuli coming from the environment surrounding the vehicle. As a result, it is possible to acquire some information: e.g., measurements of distance, light intensity, sound amplitude, detection of pedestrians, and surrounding vehicles. Therefore, measurements from exteroceptive sensors are interpreted by the vehicle to produce meaningful environmental features. 

Proprioceptive sensors, inseparable from vehicle powertrain and chassis, are widely used in vehicle production. In contrast, exteroceptive sensors are mostly used in luxury vehicles, vehicles with some level of autonomy, or experimental vehicles. Conventionally, the proprioceptive sensors are designed to measure single-process systems and are therefore limited in capacity. They are unexposed, protected from the external environment. In contrast, exteroceptive sensors are designed to analyze and monitor internal (vehicle cabin) and external environments. Thus, they are able to operate in different conditions, including with a higher degree of difficulty~\cite{Sjafrie:2020} (e.g., rain, humidity, snow, night time, etc.).

\subsubsection{Active and passive sensors}
\label{subsubsec:classification_active_passive}

Proprioceptive and exteroceptive sensors are designed to just capture and read a specific metric, or to interact with the environment by observing and recording changes in it, or reactions from it. This leads to classifying sensors as  active or passive. \textit{Passive} sensors are able to perform measurements without interacting with the environment, in other words, the sensor receives energy stimuli from the environment. \textit{Active} sensors interact with the environment to acquire data. For example, they may emit waves outside the vehicle and measure the level of the environment reaction to those waves. Wave emitters can be lasers or radars, among others.

\subsection{OTS telematics devices}
\label{subsec:OTS_devices}

While smartphones include a large number of sensors (e.g., GNSS, camera, microphone, accelerometer) which make them particularly suitable for insurance telematics~\cite{Wahlstrom:2017}, other sensors require dedicated hardware and installation process. Next, we present a background of OTS telematics devices that carry exteroceptive sensors.

\smallskip\subsubsection{\textit{OBD-II dongles and CAN bus readers}} \label{subsubsec:OBD_CAN}
A modern vehicle can contain more than one hundred sensors, generally associated with the mechanics and 
operation of the engine and vehicle systems~\cite{Fleming:2008}. Automotive systems concentrate in three areas of the vehicle: powertrain, chassis, and body. In each area, a set of sensors measures physical quantities associated to specific functions. Measurements are sent to the ECU of each system, where they are interpreted in a look-up table~\cite{Wong:2012}. Data is stored in profiles used to control the vehicle actuators and their performance, e.g. battery level, fuel injection duration, speed control, vehicle stability, anti-lock brake system, among others. The use of specific sensors may also be associated with other factors such as legislation and safety~\cite{Orazio:2011}. Data profiles from the ECUs are used to check the vehicle status information through the On-Board Diagnostics II (OBD-II) interface. It provides access to the vehicle sub-systems controlled by the ECUs, via the CAN bus. OBD-II is widely used by the automotive manufacturers for the analysis of data collected by the ECUs, and their subsequent general diagnosis. Nevertheless, the acquisition of data 
through the OBD-II connector is limited to a single port and is specific to each manufacturer which defines proprietary message codes. Commercial OBD-II dongles and more broadly CAN bus readers are connected to the power source of the vehicle itself and may have extra sensors, like a GNSS or an accelerometer.

\smallskip\subsubsection{Digital tachograph}\label{subsubsec:digital_tachograph} Commercial and utility vehicles often use an equipment to log trajectory data, such as speed and distance traveled. Digital tachograph displays, records and stores these measurements internally for the driver's work periods, which are defined by the regulatory authority. The tachograph usually uses a hall effect sensor located in the gearbox of the vehicle, or other mechanical interface whose movement is representative of the speed~\cite{Furgel:2006}. Data is stored in the driver's smart card and then mainly analyzed by fleet management tools. In fact, although tachographs provide precise and secure information, they are relegated to fleet management due to its acquisition and installation cost.

\smallskip\subsubsection{Black-box and windshield devices}\label{subssubsec:bbox_wind} 

Usually, black-box and windshield devices are installed within the vehicle and they are equipped with a self-contained sensor systems or they acquire information in a piggy-back process via the CAN bus. These devices embed a GNSS and an accelerometer sensor to define driving profiles about harsh acceleration, braking or impact. 
In addition, a windshield device may contain a SIM card and a microphone to establish a voice communication with remote assistance.

\smallskip\subsubsection{Dashcams}
\label{subsubsec:dashcam}

A dashcam is an on-board camera, usually mounted over the dashboard, that records the vehicle front view. Common uses include registering collisions, road hazards, in addition to offering video surveillance services~\cite{Kim:2017}. Since the amount of information generated by the video frames is considerable, images are selected beforehand by the processing system. Additional dashcam functionalities include gesture and voice biometric~\cite{Tymoszek:2020}.
It is worth noting that the utilization of dashcams is limited in some countries due to privacy concerns~\cite{Kim:2020:dashcam}.

\smallskip\subsubsection{Smartphones}
\label{subsubsec:smartphones} 

Smartphones involve a variety of technologies that make them a sophisticated computer, with the ability to process data and graphics, not to mention communication and sensing capabilities~\cite{Xu:2014}. Smartphones possess a large number of built-in sensors that allow continuous data collection. Added to mobility, it results in the empowerment of various types of applications with specific requirements in terms of complexity, granularity and response time. 
As for vehicular telematics, smartphones play an important role: they can acquire CAN bus data through an OBD-II dongle, a Wi-Fi or Bluetooth connection and, as such, monitor and record data from both proprioceptive and exteroceptive sensors~\cite{Wahlstrom:2017}.%

\smallskip\subsubsection{Wearable devices}
\label{subsubsec:wearable} 

Complementary to smartphones, wearable devices are used to monitor human physiological and biometric signals~\cite{Seneviratne:2017}. In the vehicular telematics context, they are used for safety and driving behavior applications~\cite{Sun:2017}. Wearable devices include smartwatches, smart glasses, smart helmets~\cite{Rajathi:2019} and electrocardiogram (ECG) sensors.

\section{Exteroceptive sensors for telematics}
\label{sec:_ext_sensors_telematics}

This section describes in detail exteroceptive sensors which are used for telematics purposes. 
Table~\ref{tab:exteroceptive-sensors-comparative} summarizes the main features of each sensor.

\begin{table*}[ht]
\caption{Multi-dimensional comparative among exteroceptive sensors used in vehicle telematics.}
\label{tab:exteroceptive-sensors-comparative}
\resizebox{\textwidth}{!}{%
\begin{tabularx}{1.0\textwidth}{>{\centering\arraybackslash}p{1.4cm}>{\centering\arraybackslash}p{0.8cm}>{\centering\arraybackslash}p{2.7cm}>{\centering\arraybackslash}p{0.9cm}>{\centering\arraybackslash}p{1.5cm}>{\centering\arraybackslash}p{4.5cm}>{\centering\arraybackslash}p{3.6cm}}
\toprule
\textbf{Sensor} &
  \textbf{Price} &
  \textbf{Main usage} &
  \textbf{Precision} &
  \textbf{Range} &
  \textbf{Advantage} &
  \textbf{Limitation}
   \\ \toprule
\multirow{2}{*}{GNSS} &
  \multirow{2}{*}{Low} &
  \multirow{2}{*}{Navigation, positioning} &
  Medium/ High & 
  \multirow{2}{*}{n/a} &
  \multirow{2}{*}{High coverage, small form factor} &
  Signal blocking in urban canyons \\ \midrule
\multirow{2}{*}{Magnetometer} &
  \multirow{2}{*}{Low} &
  Navigation, positioning, orientation &
  \multirow{2}{*}{Medium} &
  \multirow{2}{*}{n/a} &
  Small form factor, low energy consumption &
  \multirow{2}{*}{Magnetic interference} \\ \midrule
\multirow{2}{*}{Microphone} &
  \multirow{2}{*}{Low} &
  Surveillance, assistant, environmental sensing &
  \multirow{2}{*}{n/a} &
  \SI{150}{m}, omnidirectional &
  Small form factor, low energy consumption, direction of arrival &
  \multirow{2}{*}{Environmental noise} \\ \midrule
Biometric &
  Low &
  Heath monitoring &
  High &
  n/a &
  Simple data processing &
  Uncomfortable \\ \midrule
Ultrasonic &
  Low &
  Environmental sensing &
  Low\,(cm) &
  \SI{150}{cm} &
  Small form factor &
  Low resolution \\ \midrule
\multirow{2}{*}{Radar} &
  Low/ Medium &
  \multirow{2}{*}{Environmental sensing} &
  \multirow{2}{*}{High} &
  \multirow{2}{*}{\SI{250}{m}} &
  Robust in adverse climatic conditions and with scarce or absent illumination &
  Energy consumption, data processing for classification \\ \midrule
\multirow{2}{*}{LiDAR} &
  \multirow{2}{*}{High} &
  \multirow{2}{*}{Environmental sensing} &
  \multirow{2}{*}{High} &
  \SI{200}{m}, omnidirectional &
  Low sensitive to light and to weather conditions, 3D representation &
  \multirow{2}{*}{Data processing latency} \\ \midrule 
\multirow{2}{*}{Camera} &
  Medium/ High &
  \multirow{2}{*}{Environmental sensing} &
  Medium/ High &
  \multirow{2}{*}{Line-of-Sight} &
  Multiple techniques for data processing &
  Sensitive to light and weather conditions \\ \bottomrule
\end{tabularx}%
}
\end{table*}

\subsection{Global Navigation Satellite System (GNSS)}
\label{subsec:gnss}

Some OTS devices implement Location-Based Systems (LBS) using an embedded GNSS receiver. GNSS systems allow a quite accurate localization (on the meter scale) on earth, through trilateration signals from dedicated geostationary artificial satellites. Depending on the platform on which OEM devices operate, different LBS are offered. In smartphones, some location services merge short and long-range wireless networks such as Wi-Fi, Bluetooth, and cellular networks~\cite{Zandbergen:2009}, in addition to GNSS data~\cite{Dabove:2019}. 
Nowadays, Android-based devices, use messages based on the NMEA 0183 standard~\cite{NMEA:0183}. The latest updates to this standard include measurement of the pseudo-range and Doppler shift; this adds simplicity and robustness to the processing of raw GNSS measurements~\cite{GNSS:2018, Android_Dev:2019}. Nevertheless, GNSS reception exhibits outages due to interference, signal propagation, and measurement accuracy in urban canyons due to multipath effects and Non-Line-of-Sight (NLoS) conditions~\cite{Zhang:2011}. 

\subsection{Magnetometer}
\label{subsec:magnetometer}
The main function of a magnetometer is reading the strength of the Earth's magnetic field determining its orientation. Magnetometers embedded in commodity devices like smartphones have microelectromechanical systems (MEMS) which inform the magnetic field on three-axis with \SI{}{\micro\tesla} sensibility~\cite{Jones:2010}. Moreover, its miniaturized form factor and low energy consumption favors its availability in a large number of devices. Thus, it results as an important component for providing navigation and LBS services.

\subsection{Microphone}
\label{subsec:microphone}
A microphone transforms sound waves into electrical energy. These sensors are embedded as MEMS devices or condensed mics that are connected to OTS devices. Microphones are an affordable solution for real time signal processing. According to  ISO\,9613-2 standard, their sensing range reaches up to \SI{200}{m} for high intensity sounds in an urban scenario~\cite{ISO_9613-2}. Moreover, microphones consume low energy,  have very small size, and omni-directional sensing capability. Devices with an array of microphones are used to estimate the Direction of Arrival (DoA) and localize the sound source calculating the time difference of arrival between each microphone pair.
On the other hand, their efficiency largely depends on their sensitivity, sound waves amplitude, and environmental noise. 

\subsection{Biometric sensors}
\label{subsec:biometric}

Biometric sensors are used to collect measurable biological characteristics (biometric signals) from a human being, which can then be used in conjunction with biometric recognition algorithms to perform automated person identification. ECG devices installed in the steering wheel and in the driver's seat to measure heart activity, through touch or photoelectric sensors. In telematics, it is used as a proxy of drivers' stress condition, drowsiness, and fatigue~\cite{murugan2020detection}.

\subsection{Ultrasonic sensor}
\label{subsec:ultrasonic} 

Ultrasonic refers to acoustic waves, where a transmitter sends sound waves, and a receiver captures the bounce off waves from nearby objects. The distance of such object is determined through the Time-of-Flight (ToF). These waves are propagated at the speed of sound (that depends on the density of the propagation medium), and use frequencies higher than those audible by the human ear, between 20 and \SI{180}{kHz}~\cite{Siegwart:2004}. Sound propagation occurs conically, with opening angles between \ang{20} and \ang{40}. The ultrasonic sensor is suitable for low speed, short or medium range applications (tens or hundreds of cm) like parking assistance, blind spot detection and lateral moving. With a low power consumption (up to \SI{6}{W}) and a price under \SI{100}{\$}, it is a relatively affordable object detection sensor.

\subsection{Radar}
\label{subsec:radar} 

Radar (Radio Detection and Ranging) detectors use reflected electromagnetic waves. The device transmits radio wave pulses that bounce on objects outside the vehicle. The reflected pulses which arrive some time later at the sensor allow inferring different information. It is possible to determine the direction, distance, and estimate the object size~\cite{Sjafrie:2020}. The relative speed of moving targets can be calculated through frequency changes caused by the Doppler shift. Radar systems transmit waves in Ultra High Frequency (UHF), at 24, 77, and \SI{79}{GHz}, with opening angles between \SI{9}{\degree} and \SI{150}{\degree}, and elevation up to \SI{30}{\degree}. Radar can operate in distance ranges up to \SI{250}{m}, with a power consumption from \SI{12}{W}, it is used for short, mid and long-range object detection and adaptive cruise control at high speeds. Radars are robust in adverse climatic conditions (e.g., fog or rain) and with scarce or no lighting. Nevertheless, signal processing is harder for classification problems if not combined with other sensor readings. Radar's price ranges from \SI{50}{\$} to \SI{200}{\$}.

\subsection{LiDAR}
\label{subsec:lidar}

LiDAR (Light Detection And Ranging) uses laser reflection instead of radio waves.
The LiDAR sensor transmits light pulses to identify objects around the vehicle. Typically, a LiDAR emits a \SI{905}{nm} wavelength laser light beam to illuminate the objects. Pulses of laser light are generally emitted at every \SI{30}{ns}. The returning light component is coaxial with the light beam emitted by the sensor~\cite{Siegwart:2004}. The LiDAR sweeps in a circular and vertical fashion; the direction and distance of the reflected pulses are recorded as a data point. Moreover, a set of points then constitutes a point cloud which is a spatial representation of coordinates, enabling 3D model processing with high accuracy. LiDAR sensors can cover at \ang{360} the horizontal field of view around the vehicle, and up to \ang{42} the vertical field of view. LiDARs are less sensitive to light and weather conditions. Nevertheless, processing the whole LiDAR points is time consuming, thus it is not suitable for real time applications. There are low-cost LiDAR sensors (from \SI{100}{\$}) and low power consumption (from \SI{8}{W}); nonetheless, these are limited to one laser beam. More advanced models of LiDAR sensors contain laser arrays (up to 128), improving the point cloud resolution; this represents a higher energy consumption (up to \SI{60}{W}), and more expensive (up to 75,000\,\$).

\subsection{Camera}
\label{subsec:camera}

A camera is a vision sensor used to take images both inside and outside the vehicle, to detect objects on the road as well as to analyze the behavior of the driver and his environment inside the vehicle. CMOS-based cameras are widely used in vehicular applications~\cite{Miller:2004}. These can operate in the Visible (VIS) and Near-Infrared (NIR) spectral region~\cite{Siegwart:2004}. VIS cameras are largely used because these reproduce instantaneous images like those perceived by the human eye. Differently, NIR cameras detect objects based on heat radiation. Additionally, the quality of the images depends on the resolution and field of view of the device. Furthermore, vehicular applications use monocular cameras, stereo cameras, in addition to using so-called fish-eye lenses, which generate optical effects. Optical cameras are less expensive than LiDAR sensors and very effective, with power consumption less than \SI{2.5}{W}. Despite the fact that the camera generates the highest amount of data per second, accurate methods for object detection and recognition through image processing exist nowadays, like Convolutional Neural Networks (CNN) and deep learning, enabling to handle real images better than LiDAR. Some drawbacks exist though: image quality depends on lighting and weather conditions, and scene representation is limited to the pointing direction and line-of-sight.

\section{Services and Applications}
\label{sec:services_and_applications}

Next, we cover practical telematics applications and services that use exteroceptive sensors exclusively.
We organize them into four macro-areas: safety, driving behavior, road monitoring, and navigation. Given the rich literature in vehicle telematics, for each macro-area we have selected relevant works in terms of practicability, novelty, relevance, and release date. In addition, we report on available datasets if applicable. 

\subsection{Safety}
\label{subsec:safety}

As summarized in Table~\ref{tab:safety}, the safety area includes four categories of applications, related to vehicle maintenance, driving and external events: tire wear, collision detection, collision avoidance and lane departure. 

\begin{table}[t]
\caption{Safety applications with exteroceptive sensors.}
\label{tab:safety}
\begin{tabularx}{\columnwidth}{CCC}
\toprule 
\textbf{Application}                      & \textbf{Sensor/\textcolor{gray}{Dataset}} & \textbf{References} \\ \toprule
    Tire wear                            & Radar   & \cite{Matsuzaki:2012, Prabhakara:2020:osprey}                \\ \midrule
    \multirow{2}{*}{Collision detection} & Microphone & \cite{foggia_crowded_roads, Foggia2015, Foggia2016, saggese_sorenet,  morfi2018deep, 10.1145/3378184.3378186, 7472921, huang2020urban, crashzam, crashzam_springer}\\
    & \textcolor{gray}{Dataset}   & \cite{crashzam, dataset:gemmeke, dataset:mesaros, dataset:Piczak, dataset:Salamon:2014, dataset:stowell}              \\ \midrule
    \multirow{5}{*}{Collision avoidance} & Camera  & \cite{gloger2005camera, Kilicarslan:2019, Kim:2012, Wu:2018:CA}                \\
    & Radar & \cite{joshi:2019:CA, Blanc:2004, Sun:2012:CA, Park:2003}                   \\
    & LiDAR   & \cite{kumar:2019:CA, Kampker:2018, Natale:2010, ogawa:2011:CA, Nashashibi:2008}                \\
    & LiDAR+Camera  & \cite{wei2018lidar, cho2019study, Nobis:2019}          \\
    & Microphone & \cite{Mizumachi:2014}  \\ \midrule
    \multirow{3}{*}{Lane departure} & Camera & \cite{Popken:2007, Freyer:2010, Ono:2016, Grimm:2013, Andrade:2018, Gaikwad:2015, Boutteau:2013, Baili:2017}\\
    & LiDAR & \cite{Ghallabi:2018, kammel2008lidar, hata2014road, zhang2010lidar, wu2020automatic} \\
    & \textcolor{gray}{Dataset} & \cite{Aly:2008, Wu:2012, Fritsch:2013, VPGNet:2017, TuSimple:2017, CuLane:2017, Berriel:2017, ApolloScape:2019, BDD100K:2018} \\ \bottomrule                  
\end{tabularx}%
\end{table}

\smallskip\subsubsection{Tire wear}
\label{subsubsec:tire_wear}
Often underestimated, tires play a crucial role for vehicle stability and control, especially on slippery road surfaces. Nevertheless, tire wear is challenging to continuously measure due to the position and the dynamics of the tires. Matsuzaki \textit{et al.}~\cite{Matsuzaki:2012} analyze the tire surface deformation. The system uses a wireless CCD camera attached to the wheel rim to obtain 3D images. Digital Image Correlation Method (DICM) is used to estimate the strain distribution and friction load in the tire. Results show a 10\% error range in the tire load. Osprey, is a debris-resilient system designed to measure tread depth without embedding any electronics within the tire itself~\cite{Prabhakara:2020:osprey}. Instead, Osprey uses a mmWave radar, measuring the tire wear as the difference between tire tread and groove.

\smallskip\subsubsection{Collision detection}
\label{subsubsec:crash_detection}

The correlation between first aid time delay and death probability when a severe car accident occurs has been statistically proved~\cite{SnchezMangas2010ThePO}. For this reason, an automatic collision detector and rescue caller, like the eCall system\cite{eu:ecall2}, is compulsory for all the new cars sold in the European Community. For older car models devoid of eCall system, some retrofit devices are available in form of a black-box, which is connected to a \SI{12}{V} socket plug and the OBD interface~\cite{8000985, bosch:ecallretrofit, bosch:ecallconnectivity, splitsecnd:ecall}. Such devices rely on a 3-axis accelerometer to detect the collision impact. Nevertheless, when the accelerometer sensor is not firmly attached to the vehicle chassis, the acceleration measurement is not reliable. Also, the accelerometer is prone to false positives, e.g. after a street bump or if a pothole is hit and OBD dongles tend to fold out during impacts.

Recent sound event recognition breakthroughs make possible to detect a car accident through sound analysis, without a specific requirement on the microphone position or orientation. Foggia {\it et al.} were among the first to design a model for urban sound recognition, including car crashes~\cite{foggia_crowded_roads, Foggia2015, Foggia2016}. Initially, their audio classification was based on a combination of bag of words and Support Vector Machine (SVM) with features extracted from raw audio signals. Successively, Deep Neural Networks (DNN) models, in the form of CNN, have proved their effectiveness in classifying audio signals from their spectrogram representation~\cite{saggese_sorenet}. Their solutions are focused on the creation of a larger and inclusive road side surveillance system or microphone-based Intelligent Transportation Systems (ITS) like other works~\cite{morfi2018deep, 10.1145/3378184.3378186, 7472921, huang2020urban}.

Sammarco and Detyniecki~\cite{crashzam} shift the focus to driver and passengers safety, training a SVM model directly on crash sounds recorded inside the car cabin and running on a mobile application. All the other sounds supposed to be reproduced within vehicles like people talking, radio music, and engine noise is treated as negative samples, instead. The proposed Crashzam solution does not require a road side surveillance infrastructure favoring scalability. Moreover, the same authors provide a method for impact localization~\cite{crashzam_springer}. The aim is to provide a quick damage assessment and a fast querying for spare parts. It is based on a four microphones device placed at the center of the car cabin and on the knowledge of the vehicle sizes. Besides the particular dataset of events recorded within the car cabin~\cite{crashzam}, more generic urban event audio dataset exist~\cite{dataset:gemmeke, dataset:mesaros, dataset:Piczak, dataset:Salamon:2014, dataset:stowell}. These datasets contain sound clips with features extracted through the Mel-Frequency Cepstral Coefficients (MFCC) algorithm, and Machine Learning (ML)-based classification techniques. 

\smallskip\subsubsection{Collision avoidance}
\label{subsubsec:collision_avoidance}

Collision alerts warn drivers when a collision is imminent, or whether other vehicles or objects are detected extremely close. Collision Avoidance (CA) procedure includes: \textit{(i)} environment sensing for object detection, \textit{(ii)} collision trajectory and impact time estimation, \textit{(iii)} alert launching. A survey of collision avoidance techniques is provided in~\cite{Mukhtar:2015}. 
Object detection for CA involves challenges such as pedestrians, vehicles, and obstacles detection at the front as well as at the rear or sides of the vehicle.

Object detection makes extensive use of image acquisition via different kinds of cameras. Those include monochrome, RGB, IR, and NIR cameras~\cite{gloger2005camera, Kilicarslan:2019, Kim:2012, Wu:2018:CA}. A peculiarity of detecting objects by image-based sensors is the use of bounding boxes. The advantage is to crop the image around the object itself resulting in decreased computational time for post-processing. In addition to visual imaging, other active-range sensors such as radar~\cite{joshi:2019:CA, Blanc:2004, Sun:2012:CA, Park:2003}, and LiDAR~\cite{kumar:2019:CA, Kampker:2018, Natale:2010, ogawa:2011:CA, Nashashibi:2008}, can determine the proximity of objects around the vehicle in a two or three-dimensional representation. Another strategy consists of sensor fusion, for example combining camera and LiDAR~\cite{wei2018lidar, cho2019study, Nobis:2019}. Since such sensors for object detection are complementary and coordinated, they create a more resilient CA system. Surround sound can also be used for object detection. Mizumachi \textit{et al.}~\cite{Mizumachi:2014} propose a sensing method relying on a microphone array to warn drivers about another vehicle approaching from the rear side. They employ a spatial-temporal  gradient method in conjunction with a particle filter for fine DoA estimation. 

Vehicles equipped with multiple exteroceptive sensors are used to conduct experiments on various research areas and, in particular, on object detection for CA. 
Table\,\ref{tab:datasets_object_detection} lists available datasets with external perception data. These datasets are available both in images and semantically. In principle, detected objects are marked with 2D or 3D bounding boxes. Based on these markings, it is possible to categorize the detected objects, mostly of the times using neural networks. Some datasets are limited in terms of time or distance traveled (Table\,\ref{tab:datasets_object_detection}).

\begin{table}[t]
\caption{Datasets for object detection with exteroceptive sensors.}
\label{tab:datasets_object_detection}
\begin{tabularx}{\columnwidth}{>{\centering\arraybackslash}p{2.1cm}>{\centering\arraybackslash}p{0.5cm}>{\centering\arraybackslash}p{0.5cm}>{\centering\arraybackslash}p{0.6cm}>{\centering\arraybackslash}p{0.65cm}>{\centering\arraybackslash}p{0.65cm}>{\centering\arraybackslash}p{0.8cm}}
\toprule
\multirow{2}{*}{\textbf{Dataset}}                                 & \textbf{Fra\-mes} & \textbf{Sce\-nes} & \multirow{2}{*}{\textbf{Label}}    & \textbf{Annot. types} & \textbf{Annot. frames} & \multirow{2}{*}{\textbf{Size}}     \\ \toprule
KITTI~\cite{KITTI:2012}                  & 43\,k  & 22     & 3D       & 8            & 15\,k         & 1.5\,h   \\ \midrule
KAIST~\cite{KAIST:2018}                  & 95\,k  &  --    & 2D/3D    & 3            & 8.9\,k        & --     \\ \midrule
Nuscenes~\cite{Nuscenes:2019}            & 40\,k  & 1\,k   & 2D/3D    & 23           & 40\,k         & 5.5\,h   \\ \midrule
DDAD~\cite{DDAD:2020}                    & 21\,k  & 435    & 2D/3D    &      --      & 99\,k         &     --   \\ \midrule
A2D2~\cite{A2D2:2019}                    & 41\,k  &   --    & 3D & 38           & 12\,k         &     --   \\ \midrule
ApolloScape\cite{ApolloScape:2019}      & 144\,k & 103    & 2D/3D    & 28           & 144\,k        & 100\,h   \\ \midrule
BDD100K~\cite{BDD100K:2018}            & 100\,k &   --   & 2D       & 10           & 100\,k        & 1,000\,h  \\ \midrule
Waymo~\cite{Waymo:2019}                  & 12\,M  & 1.1\,k   & 2D/3D    & 4            & 230\,k        & 6.5\,h   \\ \midrule
Vistas~\cite{Vistas:2017}      & 25\,k  &   --   & 2D       & 66           & 25\,k         & 6.5\,h   \\ \midrule
Cityscapes~\cite{Cityscapes:2016}        & 25\,k  &   --   & 2D       & 30           & 25\,k         &   --    \\ \midrule
\multirow{2}{*}{Argoverse~\cite{Argoverse:2019}} &   \multirow{2}{*}{--}   & 113    & 3D       & \multirow{2}{*}{15}           & \multirow{2}{*}{22\,k}         & 1\,h     \\
&        & 324\,k   & 2D       &              &               & 320\,h   \\ \midrule
H3D~\cite{H3D:2019}                      & 27\,k  & 160    & 3D       & 8            & 27\,k         & 0.77\,h  \\ \midrule
Oxford~\cite{Oxford:2019}       &   --   & 100\,+ & 2D/3D    &     --      &      --     & 1,000\,km \\ \midrule
Eurocity~\cite{Eurocity:2019}            & 47k    &   --   & 2D       & 8            &    --         & 53\,h    \\ \midrule
Canadian~\cite{Canadian:2020}            & 7k     &   --   & 2D/3D    & 16           &    --         &   --     \\ \midrule
Lyft5~\cite{Lyft_Prediction:2020}        &  --    & 170\,k & 3D       & 9            & 46\,k         & 1,118\,h  \\ \midrule
D$^2$-City~\cite{D2City:2019}            & 700\,k &    --  & 2D       & 12           &   --          & 55\,h    \\ \midrule
BLVD~\cite{BLVD:2019}                    & 120\,k &   --   & 2D/3D    & 28           & 250\,k        &   --     \\ \midrule
Honda~\cite{Honda:2018}                  &   --   &   --   & 2D       & 30           &   --          & 104\,h   \\ \midrule
Ford AV~\cite{Ford:2020}                 &   --   &   --   & 3D       &    --        &   --          & 66\,km  \\ \midrule
Astyx~\cite{Astyx:2019}                  & 546    &  --    & 2D/3D    &     --       &    --         &   --     \\ \bottomrule
\end{tabularx}
\end{table}

\smallskip\subsubsection{Lane departure}
\label{subsubsec:Lane_departure}

Lane detection (LD) and tracking (LT) is a hot topic in the driving safety area due to the complexity needed to achieve reliable results. Most of the applications using LD and LT aim to warn the driver about an odd trajectory before lane crossing to prevent accidents. The first challenge is to correctly extract lane from the acquired image in a single-frame context. This process must be both quick and precise. 

Automotive manufacturers deploy inexpensive cameras, usually on the vehicle windshield. Audi implements a monochrome camera with a CMOS image sensor. When the driver performs any maneuver that is considered dangerous by the system, a vibration of the steering wheel is produced~\cite{Popken:2007, Freyer:2010}. Toyota uses monocular cameras in the vehicles to detect LD and send alerts for lane-keeping. These cameras are equipped with a single lens for detecting white lane markings and headlights~\cite{Ono:2016}. Mercedes-Benz uses a stereo camera for its lane-keeping system. Starting from the detection of lane markings, a steering assistant interacts with the driver to facilitate vehicle driving. The system also uses vibration of the steering wheel to alert the driver~\cite{Grimm:2013}.

Andrade \textit{et al.}~\cite{Andrade:2018} propose a three-level image processing strategy for LD. In the low-level, the system essentially performs image compression and delimits the Region of Interest (ROI). In the mid-level, it uses filters to extract features. Finally, high-level processing uses the Hough transform algorithm to extract possible line segments in the ROI. A similar approach is employed by Baili \textit{et al.}~\cite{Baili:2017}. With the same LD technique, Gaikwad and Lokhande~\cite{Gaikwad:2015} use a Piecewise Linear Stretching Function (PLSF) in combination with the Euclidean distance transform to keep false alarms under 3\% and the lane detection rate above 97\%.  
The PLSF converts images to grayscale in binary mode and improves contrast in ROI. Boutteau \textit{et al.}~\cite{Boutteau:2013} employ fish-eye cameras to detect lane lines, and from projecting lines onto a unitary virtual sphere, triangulate its projection in perspective, reconstructing the road lines in 3D. Omnidirectional line estimation uses the RANSAC (RANdom SAmple Consensus) method. The system has a true positive rate of 86.9\%. As road markings are reflective, they can be detected using intensity laser data coming from a LiDAR~\cite{Ghallabi:2018, kammel2008lidar, hata2014road, zhang2010lidar, wu2020automatic}. Following this intuition, LD includes road segmentation detecting curbs with elevation information and selecting the most reflective points from the road plane.

To study lane detection, there are some datasets available with vision-based real data. These have been collected under different climatic and light exposure conditions. The datasets consist of video sequences, images and frames of real scenes on the road~\cite{Aly:2008, Wu:2012, Fritsch:2013, VPGNet:2017, TuSimple:2017, CuLane:2017, Berriel:2017, ApolloScape:2019, BDD100K:2018}. 

\subsection{Driving behavior}
\label{subsec:driving_behavior}

One of the main risk factors on the roads is the human driving~\cite{Singh:2015}. The driver behavior is associated with different events that generate dangerous or aggressive actions. As shown in Table\,\ref{tab:driving_behavior}, driving behavior can be evaluated in macro-areas that study different events in driving practice. The literature on the classification of driving behaviors is rich. In the insurance market, commercial products like Pay As You Drive (PAYD) or Pay How You Drive (PHYD) determine their price taking driving behavior metrics into account~\cite{bordoff:2010}.

\begin{table}[!t]
\caption{Driving behavior applications with exteroceptive sensors.}
\label{tab:driving_behavior}
\begin{tabularx}{\columnwidth}{CCC}
\toprule \textbf{Application}                        & \textbf{Sensor/\textcolor{gray}{Dataset}} & \textbf{References}\\ \toprule
    \multirow{5}{*}{Driving profiling}     & GNSS  & \cite{Abdelrahman2019, zheng2015trajectory, Dong:2017,Dong:2016, andrieu2012comparing, chen2018driver}                  \\
    & Microphone  & \cite{Goksu:2018, Kubera:2019, Ma:2017}             \\
    & Camera  & \cite{Tran:2012}                 \\
    & Tachograph  & \cite{Rygula:2009, Kim:2016, Zhou:2019}            \\ 
    & \textcolor{gray}{Dataset} & \cite{Hankey:2016, dataset:campbell2012shrp,  dataset:VTT1/LYUBJP_2020, dataset:VTT1/KVCO0B_2017, LeBlanc:2010, Bender:2015, site:nhtsa} \\ \midrule
    Driver detection                       & Microphone  & \cite{chu:2014, Yang:2011}            \\ \midrule
    \multirow{2}{*}{Driver identification} & GNSS  & \cite{Jafarnejad:2019}                  \\ 
    & \textcolor{gray}{Dataset} & \cite{Abut:2007}                \\
    \midrule
    \multirow{3}{*}{Driver health monitoring}     & Plethysmograph   & \cite{Shin:2010}        \\
    & ECG  & \cite{Cassani:2019, Jung:2014, Wartzek:2011, Sakai:2013}                   \\
    & Biometric & \cite{Sinnapolu:2018,  Audi:2016,  MB:2019,  Nismo:2013}              \\ \midrule
    \multirow{5}{*}{Driver distraction} & Camera & \cite{Fridman:2016, Tran:2018, Zhang:2019, Walger:2014, Wijnands:2019, Hossain:2018, Xu:2014:soberDrive, Chuang:2014, Qiao:2016, You:2013, abouelnaga2017realtime} \\
    & Microphone  & \cite{Xie:2019, Xu:2017}           \\
    & ECG   & \cite{BenDkhil:2015, Yeo:2009}                  \\
    & Infrared  &  \cite{Bhaskar:2017, Lee:2008}               \\
    & \textcolor{gray}{Dataset}  & \cite{Taamneh:2017}     \\ \bottomrule        
\end{tabularx}%
\end{table}

\smallskip\subsubsection{Driving profiling}
\label{subsubsec:driving_profiling}

Risk predictions are based on driving behavior and profiling, always supported by historical GNSS data, sometimes enriched with weather and traffic conditions~\cite{Abdelrahman2019, zheng2015trajectory}. Recently, Dong \textit{et al.}~\cite{Dong:2017} propose an Autoencoder Regularized deep neural Network (ARNet) and a trip encoding framework called trip2vec to learn drivers' driving styles directly from GPS records. This method achieves an identification accuracy higher than their own previous work based on different DNN architectures on characterizing driving styles~\cite{Dong:2016}. Authors in~\cite{andrieu2012comparing, chen2018driver}, instead, consider and evaluate fuel consumption and eco-driving as a proxy of driving behavior.

The identification of driving characteristics is relevant to describe different profiles of driving behavior. 
Instead of relying on low frequency GNSS points, G\"oksu~\cite{Goksu:2018} proposes to monitor the vehicle speed through acoustic signals. He employs a Wavelet Packet Analysis (WPA) for processing the engine speed variation sound. This method provides arbitrary time-frequency resolution. Given that WPA output is a sub-signals set, the author uses norm entropy, log energy and energy. These features feed a Multi-Layer Perceptron (MLP). Experiments were conducted with data was collected from four different vehicles, using a digital recorder attached to a microphone, located at \SI{1}{m} away from the engine. Best results are obtained with the norm entropy as feature. On the other hand, Kubera \textit{et al.}~\cite{Kubera:2019} study the drivers' behavior approaching speed check points recording and analyzing audio signals recorded by a roadside microphone. They test multiple ML models (SVM, random forest, and ANN), as well as a time series-based approach to classify car accelerating, decelerating, or maintaining constant speed. Results shows 95\% classification accuracy in speed estimation. Microphone is also used for detecting turn signals in a larger framework for auto-calibrating and smartphone-based dangerous driving behavior identification system~\cite{Ma:2017}.

Assessments of data collected in commercial vehicles through tachographs are performed to analyze the behavior of drivers and detect dangerous events in shared driving. Data collected is of particular interest to the vehicle owner and manufacturer, telematics insurance, and a regulatory entity. Data collected through tachographs loaded on commercial vehicles are studied in~\cite{Rygula:2009, Kim:2016, Zhou:2019}. Also, camera is used for modeling and predicting driver behavior~\cite{Tran:2012}. The camera is actually pointed to drivers' feet to track and analyze their movements through a Hidden Markov Model (HMM). The model is able to correctly predict brake and acceleration pedal presses 74\% of time and \SI{133}{ms} before the actual press. Instrumenting vehicles with data recorders and transmitters to collect data to study and assess driving behavior is expensive and often have to face privacy issues. 
Nevertheless, many academic institutions and public authorities have built, on voluntary-basis, databases of real (also called ``naturalistic'') rides~\cite{Hankey:2016, dataset:campbell2012shrp,  dataset:VTT1/LYUBJP_2020, dataset:VTT1/KVCO0B_2017, LeBlanc:2010, Bender:2015} or test rides in a controlled environment~\cite{site:nhtsa}.

\smallskip\subsubsection{Driver detection}
\label{subsubsec:driver_passenger_identification}

One of the basic problems driving behavior monitoring systems is to differentiate the driver from passengers. Systems proposed to resolve this issue are named Driver Detection Systems (DDS). The DDS is a building block for mobility services, especially common for PHYD or PAYD smartphone-based applications or fleet management, as the same person can sometimes drive its own vehicle or be just a passenger on other occasions (e.g., on taxis, buses, and friends' car). 
One common approach is to split the vehicle seats in four quadrants (front/rear and left/right) where the driver occupies either front/left or the front/right seat and to analyze signals during maneuvres~\cite{chu:2014}.
Following this approach, a microphone-based solution has been proposed by Yang \textit{et al.}~\cite{Yang:2011}: supposing that the vehicle has four loudspeakers at the four corners of the car cabin, and the driver/passenger's smartphone can establish a Bluetooth connection to the car's stereo system, then some high frequency beeps are played at some predefined time intervals. On the other side, beeps are recorded and analyzed by a mobile app to deduce the reception timing difference between left/right and front/left. Despite being an elegant solution, locations equidistant from the loudspeakers present high incertitude. Moreover, Bluetooth association is often guaranteed only to the driver's smartphone.

\smallskip\subsubsection{Driver identification}
\label{subsubsec:driver_identification}

A slightly different problem is Driver Identification (DI): given a set of drivers, recognizing who is currently driving. Traditional authentication methods which use smart-cards, RFID tags, or code dialing, require the installation of specific hardware. Jafarnejad \textit{et al.}~\cite{Jafarnejad:2019} propose a DI approach based on noisy and low rate location data provided by GNSS embedded in smartphones, car-navigation system or external receivers. The authors extract characteristics through a semantic categorization of data and metrics detected in the dataset. For that, a DNN architecture analyzes the characteristics. Results show that the approach achieves an accuracy of 81\% for 5 drivers. Nonetheless, the amount of data trained can generate errors. Hence, it is necessary to implement one more authentication method in the algorithm. Moreira-Matias and Farah~\cite{Moreira:2017} propose a methodology to DI using historical trip-based data. The system uses data acquired through a data recorder, and uses ML techniques for feature analysis. Results show a high accuracy in predicting the driver category ($\approx$88\%). 

\smallskip\subsubsection{Driver health monitoring}
\label{subsubsec:Health_monitoring}

Drivers' health describes different driving actions including maneuvres leading to accidents. For this reason, researchers and industry are interested in the use of sensors for drivers' physiological electrocardiogram (ECG) signals. The correlation between ECG signals and both heart and breathing rhythms describes alterations in the body due to stress, fatigue, drowsiness, sleepiness, inattention, drunkenness, decision errors, and health issues~\cite{Choi:2016}. One strategy for obtaining ECG signals is to use resistive sensors on the steering wheel. 
Osaka~\cite{Osaka:2012} and Shin \textit{et al.}~\cite{Shin:2010} have developed a heart rate verification system using electrodes and a photo-plethysmograph, a heart-rate sensor that uses a photoelectric pulse wave.

Cassani \textit{et al.}~\cite{Cassani:2019} evaluate ECG signals through electrodes installed on the steering wheel to study three factors: ECG signal quality, estimated heart and breathing rate. Jung \textit{et al.}~\cite{Jung:2014} propose a real-time driver health monitoring system with sleepiness alerts. Another technique is the use of sensors on the back of the driver's seat. Wartzek \textit{et al.}~\cite{Wartzek:2011} propose a reliable analysis of sensors distributed according to the morphology of the driver in the back of the seat. The authors show that 86\% of the samples are reliable on a testbed of 59 people. Sakai \textit{et al.}~\cite{Sakai:2013} use resistive sensors mounted on the back of the driver's seat to analyze the heart-rate in different driving conditions, with speed variations. 

Very recently, smartwatches are becoming more and more pervasive for health and wellness monitoring~\cite{Reeder:2016}. Sinnapolu \textit{et al.}~\cite{Sinnapolu:2018} propose to monitor the driver's heart rate via smartwatch and in case of critical conditions (the driver is not responding for in-vehicle button press or driver related activity), then a micro-controller sends CAN messages to activate the auto pilot, to pull over for assistance, and route the vehicle to the closest health center.

Automakers also are careful about their customers' well-being while driving. Nissan NISMO uses wearable technology to capture biometric data from the driver via the heart rate monitor. This device can analyze heart rate, brain activity, and skin temperature. An application can determine early fatigue, concentration, emotions, and hydration level~\cite{Nismo:2013}. Audi uses smartwatches or fitness wristbands that monitor heart rate and skin temperature. The goal is to reduce stress levels while driving, besides improving the concentration and fitness driving of drivers. Audi plans a driver assistance service that can perform autonomous driving functions, assisted emergency stops, and implement emergency services via eCall~\cite{Audi:2016}. Mercedes-Benz monitors health and fitness levels. The smartwatch can transmit the data from the sensors through the smartphone and displays the data on the on-board computer. The system analyzes vital data such as stress level and sleeping quality, activating different programs depending on the individual profile of the driver~\cite{MB:2019}.

\smallskip\subsubsection{Driver distraction}
\label{subsubsec:Driving_distraction}

A recurring problem in driving behavior is distraction. In fact, just in 2017, distracted driving claimed 3,166 lives in the United States~\cite{NHTSADistractedDriving:2019}. The following are common types of distraction~\cite{NHTSADistractedDriving:2010}: 
\begin{itemize}
    \item {\it visual}: taking the eyes off the road; 
    \item {\it manual}: taking the hands off the steering wheel;
    \item {\it cognitive}: taking the mind off of driving.
\end{itemize}

Most of the existing solutions use image analysis to track drivers' eyes and gestures, whether the gaze is on the street ahead or if looking elsewhere for distraction (e.g., looking the smartphone) or for drowsiness~\cite{Fridman:2016, Tran:2018, Zhang:2019, Walger:2014, Wijnands:2019, Hossain:2018, Xu:2014:soberDrive, Chuang:2014, Qiao:2016, You:2013, abouelnaga2017realtime}. Usually, images from drivers' smartphone frontal camera are analyzed through quite complex CNN architectures for face, eyes, nodding and yawning detection and a warning sound is reproduced in case of danger. Special precautions must adopt at night when ambient lighting is scarce.

Besides images or videos analysis, Xie \textit{et al.}~\cite{Xie:2019} propose a driver's distraction model based on the audio signal acquired by smartphone microphone. They design a Long Short-Term Memory (LSTM) network accepting audio features extracted with under-sampling technique and Fast Fourier Transform (FFT). With the goal of an early detection of drowsy driving, they achieve an average total accuracy of 93.31\%. Also, Xu \textit{et al.}~\cite{Xu:2017} rely on sound acquired by drivers' smartphones to assess inattentive driving. Through an experimental campaign, they claim a 94.8\% model accuracy in recognizing events like fetching forward, picking up drops, turning back, eating and drinking. Such events exhibit unique patterns on Doppler profiles of audio signals.

As drowsiness drops the attention level, it is also considered a form of distraction. Relying on different sources of information, authors in~\cite{BenDkhil:2015, Yeo:2009} evaluate drowsiness by analysis of electroencephalography (EEG) signals records. Bhaskar~\cite{Bhaskar:2017}, instead, proposes EyeAwake, which monitors eye blinking rate, unnatural head nodding/swaying, breathing rate and heart rate to detect drowsy driving leveraging  infrared sensors consisting of an infrared Light Emitting Diode (LED) and an infrared photo-transistor. Its 70\% accuracy though does not make it attractive despite its low cost. Nonetheless, infrared sensor is used by Lee \textit{et al.}~\cite{Lee:2008} to monitor driver's head movement to detect drowsiness with 78\% of accuracy rate. Car manufacturers are very careful to drivers' fatigue and drowsiness proposing specific systems in high range products. A list of current research and market solutions is provided in~\cite{Doudou:2020}. Taamneh \textit{et al.}~\cite{Taamneh:2017} provide to the research community a multi-modal dataset for various forms of distracted driving including images, Electro-Dermal Activity (EDA) and adrenergic sensor (for heart and breathing rate) data.

\subsection{Road Monitoring}
\label{subsec:road_monitoring}

Poor road conditions produce mechanical damages, increase vehicle maintenance expenses, poor water draining, not to mention higher accident risk. Different approaches have been developed to monitor the road surface, share this information and alert drivers, as shown in Table\,\ref{tab:road_monitoring}. As smartphones possess a three-axis accelerometer, it is possible to process the vertical acceleration signal within a mobile application to find pavement asperities or to identify dangerous zones. Naricell and TrafficSense are two example applications~\cite{Nericell:2008, TrafficSense:2008}. Nevertheless, to provide reliable measures, a proprioceptive sensor like the accelerometer must be well fixed on the vehicle chassis. On the other hand, exteroceptive sensors can easily be used to overcome this limitation and to go beyond the mere pothole detection.

\smallskip\subsubsection{Road porosity}
\label{subsubsec:road_porosity}
Another way to infer road conditions is through acoustic analysis of the tire and road surface. Crocker \textit{et al.}~\cite{Crocker:2005} study the impact between tire tread, road, and air pumping with the ISO 11819\-2:2017 close-proximity (CPX) method~\cite{ISO_11819}. Through their experiments, they are able to identify surfaces which have a greater sound absorption, like porous road pavement. Such surfaces have the advantage that they drain water well and reduce the splash up behind vehicles during heavy rainfalls. In a similar context, Bezemer-Krijnen \textit{et al.}~\cite{Krijnen:2016} study the tire-road rolling noise with the CPX approach to figure out the pavement roughness and porosity, as well as the influence of tire tread and road characteristic on the noise radiation.
\smallskip\subsubsection{Road wetness}
\label{subsubsec:road_wetness}
Abdi\'c \textit{et al.}~\cite{Abdic:2016} suggest the use of a DNN model, namely a Bi-directional LSTM (BLSTM) Recurrent Neural Network (RNN) model, to detect the road surface wetness. Data is collected with a shotgun microphone placed very close to the rear tire. Authors conduct experiments at different speeds, types of road, and road International Roughness Indexes (IRI). Their model achieves an Unweighted Average Recall (UAR) of 93.2\% for all vehicle speeds. Alonso \textit{et al.}~\cite{Alonso:2014} propose an asphalt status classification system based on real-time acoustic analysis of tire-road interaction noise. Similar to~\cite{Abdic:2016}, their goal is to detect road weather conditions with an on-board system. The authors use a SVM model in combination with feature engineering extraction methods. Results show that wet asphalt is detected 100\% of the time, even using just one feature. Meantime, dry asphalt detection achieves 88\% accuracy. Yamada \textit{et al.}~\cite{Yamada:2003} study the road surface condition wetness based on images taken by a TV camera inside the vehicle. They employ light polarization techniques to distinguish between a dry surface and a surface wet of rain or snow. Jokela \textit{et al.}~\cite{Jokela:2009} also present IcOR, a method to monitor road conditions based on light polarization reflected from the road surface. To estimate the contrast of the images, the system evaluates the graininess and the blurriness of the images. IcOR uses a monochrome stereo camera pair.

\begin{table}[t]
\centering
\caption{Road monitoring applications with exteroceptive sensors.}
\label{tab:road_monitoring}
\begin{tabularx}{\columnwidth}{>{\centering\arraybackslash}p{3cm}CC}
\toprule
\textbf{Application}     & \textbf{Sensor/\textcolor{gray}{Dataset}} & \textbf{References} \\ \toprule
        Road porosity           & Microphone    & \cite{Crocker:2005, Krijnen:2016}          \\ \midrule
        \multirow{2}{*}{Road wetness}             & Microphone   & \cite{Abdic:2016, Alonso:2014}           \\
        & Camera   & \cite{Yamada:2003, Jokela:2009}               \\ \midrule
        \multirow{6}{*}{Pothole detection}        & Microphone    & \cite{Mednis:2010}          \\
        & Camera  & \cite{Hou:2007, Chun:2019, ye2019convolutional, Maeda_2018, shim2019road, anand2018crack, Huidrom:2013} \\
        & Camera+Laser   & \cite{7084929}         \\
        & Radar & \cite{Huston:2000}                   \\
        & Ultrasonic   & \cite{Madli:2015}           \\
        & \textcolor{gray}{Dataset} & \cite{road_damage_dataset_2018, kaggle_pothole_image_dataset, shi2016automatic, yang2019feature, mei2020densely}\\ \midrule
        Road slipperiness        & Tachograph  & \cite{Jang:2019}            \\ \midrule
        Road type classification & Ultrasonic  & \cite{Bystrov:2016}   \\  \midrule
        \multirow{4}{*}{Parking lots detection}  & Ultrasonic & \cite{Park:2008:parking} \\
        & Radar & \cite{Loeffler:2015} \\
        & Camera & \cite{grassi:2015:parking, grassi2017parkmaster, Ng:2017:parking} \\
        & LiDAR & \cite{Park:2019:parking} \\ \bottomrule
\end{tabularx}%
\end{table}

\smallskip\subsubsection{Pothole detection}
\label{subsubsec:pothole_detection}
Another approach of analyzing road conditions is to detect potholes and gaps. Mednis \textit{et al.}~\cite{Mednis:2010} introduce a method for pothole detection and localization called RoadMic. Their dataset includes a combination of timestamped sound fragments with GPS positions. The sound signal is low passed to discard the noise (associated with high frequencies) and to reduce transmission latency. The proposal is tested in an urban scenario, considering 10 test drives, and data is analyzed offline. Although pothole detection is based on a simple sound signal amplitude threshold and position on the triangulation of several GPS points, authors conclude that RoadMic detects potholes with more than 80\% reliability, depending on the GPS capabilities and driving speed. Hou \textit{et al.}~\cite{Hou:2007} perform pothole recognition from 2D images taken by more cameras, and with a stereovision technique, interpolate each pair of images to generate a 3D image. Their initial goal is to have a 3D reconstruction of the pavement with an accuracy of 5\,mm at vertical direction. Chun \textit{et al.}~\cite{Chun:2019} analyze road surface damage through cameras installed on the vehicle, taking photos up to \SI{100}{km/h}. The authors use a CNN to classify the images and detect surface damages. Other studies also use different CNN architectures to classify road potholes and cracks~\cite{Maeda_2018, shim2019road, anand2018crack, ye2019convolutional}. Huidrom \textit{et al.}~\cite{Huidrom:2013} quantify potholes, cracks and patches using image processing techniques supported by heuristically derived decision logic. The testbed uses a portable digital camera and a monochromatic camera.

Other sensors are also used instead of cameras for road conditions assessment. Vupparaboina~\textit{et al.}~\cite{7084929} instead, couple camera and laser scanning within a physics-based geometric framework to identify dry and wet pothole. The system analyzes the deformations of the laser light through the camera. Huston \textit{et al.}~\cite{Huston:2000} use a Ground Penetrating Radar (GPR) in the frequency band of \SI{0.05}{GHz} to \SI{6}{GHz} to analyze concrete roadways subjected to mechanical stress, especially detecting delamination conditions with signal processing. In laboratory tests, the proposed solution is able to detect defects as small as \SI{1}{mm}. Madli \textit{et al.}~\cite{Madli:2015} also use an ultrasonic sensor to identify potholes and humps as well as their depth and height respectively. As each pothole is geotagged too, the system uses a smartphone application to alert drivers approaching dangerous zones.

Other methods to detect potholes and cracks include 2D image and 3D surface analysis. 
Recent progress in image processing brought by CNNs and the presence of high resolution cameras on smartphones, make such methods very convenient and accurate. The availability of open source image datasets contribute even more to the popularity of image processing for pothole detection. Moreover, datasets are necessary to train large DNNs~\cite{road_damage_dataset_2018, kaggle_pothole_image_dataset, shi2016automatic, yang2019feature, mei2020densely}. Unfortunately, the effectiveness of smartphone camera image processing for road surface monitoring drastically drops down with poor lighting conditions and dense traffic situations.

\smallskip\subsubsection{Road slipperiness}
\label{subsubsec:road_slipperiness}
Slippery road conditions are a crucial issue for drivers. Jang~\cite{Jang:2019} identifies slippery road spots using data from digital tachographs on-board commercial vehicles. The system measures the differences between the angular and rotational speed of the wheels, calculates the linear regression of the data, and estimates the road slipperiness within the calculated confidence interval. Experiments are conducted in different surfaces and states. Results show\,$\pm$20\% of wheel slips with a 99.7\% confidence interval. Nonetheless, the system has some issues concerning GPS interference, and other strategies can depend from readings unrelated to tachograph. Therefore, the authors suggest merging the proposed method with other techniques to improve it. 

\smallskip\subsubsection{Road type classification}
\label{subsubsec:road_type_classification}
Bystrov \textit{et al.}~\cite{Bystrov:2016} investigate the use of a short-range ultrasonic sensing system to classify road surfaces: asphalt, mastic asphalt, grass, gravel, and dirt road. Among the classification methods used, MLP shows the best performance. Mukherjee and Pandey~\cite{Mukherjee:2017} classify road surfaces through the texture characterization. The authors use Gray-Level Co-occurrence Matrix (GLCM) to the texture analysis. The system uses a linearly scanning method to evaluate the GLCM approach. To test the approach, marks are introduced using a vision dataset~\cite{KITTI:2012}. The authors conclude that this tool can be added in road segmentation processes.

\smallskip\subsubsection{Parking space detection}
\label{subsubsec:parking_space_detection}
Road monitoring services also include the detection of free parking lots. In large cities, the quest for a free parking space is a time consuming and stressful task which impact driving behavior and fuel consumption. The problem of detecting free parking lots while driving and without instrumenting or changing the road infrastructure, has been initially tackled with ultrasonic~\cite{Park:2008:parking} and radar sensors~\cite{Loeffler:2015}. Successively, due to the recent advances in image object detection with computer vision and CNN, camera~\cite{grassi:2015:parking, grassi2017parkmaster, Ng:2017:parking} or LiDAR based systems have been proposed~\cite{Park:2019:parking}. The common strategy is to sense the roadside while driving and compare its occupancy with a pre-defined parking lots map. Considering all lots as free, the occupancy information is shared and vice-versa.

\subsection{Navigation}
\label{subsec:navigation}

LBS are widely used in vehicle telematics to track vehicle navigation and to guide drivers from origin to destination. Currently, the automotive sector represents 55\% of the LBS market, or 93.3\% when combined with consumer solutions~\cite{GNSSReport:2019}. Most LBS for vehicle telematics are based on GNSS receivers. Nonetheless, GNSS outages reduce the accuracy of vehicle positioning. To encompass these issues, there are employed standalone devices with embedded MEMS sensors like accelerometer and gyroscope~\cite{Tiliakos:2013}, which make up a 6-degree-of-freedom system recognized as the Inertial Measurement Unit (IMU) \cite{Seel:2014}. Through the processing of these signals, it is possible to enable the tracking, position, and orientation of the vehicle, making it an Inertial Navigation System (INS). Compared to GNSS issues, INS exhibits cumulative growth in bias sensor error~\cite{Woodman:2007, Ramanandan:2012, Prikhodko:2018}. 

Likewise, with the implementation of exteroceptive sensors in vehicles, vehicular telematics now can interact with the vehicle surroundings. In addition to determining the location of the vehicle in a map, it is possible to recognize the position relative to static or moving objects, which makes navigation systems in the vehicle more intuitive. Moreover, having information on the trajectories of the vehicle and surrounding objects, we can gather safety-related information, insurance-relevant data, as well as to build driving analysis tools to describe driver behaviors and their relevance with respect to partial or autonomous driving assistance systems. Leveraging on the functionality of the exteroceptive sensors, various works study and analyze LBS through the integration of GNSS and INS or Simultaneous Localization and Mapping (SLAM)-based systems. As shown in Table\,\ref{tab:navigation}, we consider GNSS/INS-based and SLAM-based applications, since these are widely used for vehicular navigation systems.

\begin{table}[t]
\centering
\caption{Navigation applications with exteroceptive sensors.}
\label{tab:navigation}
\begin{tabularx}{\columnwidth}{CCC}
\toprule
\textbf{Application}     & \textbf{Sensor/\textcolor{gray}{Dataset}} & \textbf{References}\\ \toprule
        \multirow{3}{*}{GNSS/INS-based}           & Camera & \cite{Schreiber:2016, Ramezani:2018, Wen:2019, Shunsuke:2015}             \\
        & LiDAR & \cite{Hata:2016, Meng:2017, Wan:2018, Demir:2019}             \\
        & Radar   & \cite{Abosekeen:2019}              \\ \midrule
        \multirow{3}{*}{SLAM-based}        & Camera  & \cite{Lemaire:2007, Magnabosco:2013, Chiang:2020}            \\
        & LiDAR  & \cite{Ghallabi:2018, Javanmardi:2019, Choi:2014, Moras:2010}\\
        & Radar  & \cite{Jose:2005, Cornick:2016, Ort:2020}          \\ \bottomrule        
\end{tabularx}%
\end{table}

\smallskip\subsubsection{GNSS/INS-based navigation}\label{subsubsec:GNSS-based} To mitigate problems with inaccuracies and sensor biases, GNSS and INS systems are used simultaneously as real-time calibration systems along with exteroceptive sensors to reduce cumulative error and the effects of GNSS outages~\cite{Zhang:2011, Ramanandan:2012, Sasani:2015}.

Schreiber \textit{et al.}~\cite{Schreiber:2016} propose a localization method using real-time camera images coupled to a GNSS/INS when GNSS measurements have low precision. The system analyzes the changes in camera orientation between pairs of frames to estimate the position and speed change. The authors implement an Extended Kalman Filter (EKF) to estimate movement through cameras and GNSS/INS prediction. Nonetheless, the system depends on the quality of the GNSS signal. Ramezani \textit{et al.}~\cite{Ramezani:2018} support stereo cameras with inertial sensors in a Visual-Inertial Odometry (VIO) system. The idea is to keep navigation operational in the absence of GNSS signal. The system implements a Multi-State Constraint Kalman Filter (MSCKF) to integrate INS data and images from a single camera. A second camera is used to impose additional conditions to improve the estimation of the system. Results show that the MSCKF stereo achieves a lower average positioning error in relation to the mono approach and the integrated with INS. 

Wen \textit{et al.}~\cite{Wen:2019} use a fish-eye camera pointing the sky to classify measurements in LoS or NLoS environments, beforehand to integrating them into GNSS/INS. The authors formulate an integration problem using the measurements of each system independently. Variable analysis results in a non-linear optimization problem, where the sensor measurements are interpreted as edges, and the different states as nodes; these are defined in a factor graph. The experiments are carried out in an urban environment. Compared with an EKF and a factor graph for the GNSS/INS system, the fish-eye camera with factor graph technique reduces the mean positioning error from \SI{8.31}{m} to \SI{3.21}{m} and from \SI{7.28}{m} to \SI{4.73}{m} in the selected scenarios. Even if the remaining positioning error is lower, yet it is too much for autonomous driving vehicles and it has not been verified with a very large experimental campaign. Shunsuke \textit{et al.}~\cite{Shunsuke:2015} present a system for locating the vehicle through positioning in the lane of a road. The system integrates a monocular camera with GNSS/INS. Analysis of the lane detection image is carried out by implementing the Inverse Perspective Mapping (IPM) algorithm, which projects the ROI onto a ground plane, and processed through the Hough transform. GNSS/INS/lane detection images use a particle filter. Results show that the average positioning error is lower than 0.8\%, and the correct lane rate is higher than 93\%. 

A drawback of image sensors is their sensitivity to both very intense or scarce lighting~\cite{Hata:2016}. Yet, an important asset for navigation is dynamic object detection. LiDARs can also serve this purpose. Hata \textit{et al.}~\cite{Hata:2016} present a vehicle location method that includes the detection of curbs and road markings through LiDAR readings. Curb detection is based on the adjacent distance between rings formed by the sensor readings; a gradient filter analyzes the false classification of curbs, and a regression filter adjusts a function to remove outliers and to consider candidate points. For detecting road markings, the authors use the Otsu threshold method, an algorithm that returns a single intensity level in a pixel, and a reflective intensive sensor calibration method. The binary map grid for both curbs and road markings is integrated with GNSS/INS using the Monte Carlo Location (MCL) algorithm, a method that estimates the position by matching the sensor measurements and the area where the vehicle displaces. Results show that the longitudinal and lateral errors are less than \SI{0.3}{m}. 

Meng \textit{et al.}~\cite{Meng:2017} propose a vehicle location system based on GNSS, IMU, Distance-Measuring Instrument (DMI), and LiDAR. GNSS/IMU/DMI systems are combined employing a fault-detection method based on Unscented Kalman Filter (UKF) and a curb detection. The system calculates the lateral location of the vehicle and estimates the lateral error. Wan \textit{et al.}~\cite{Wan:2018} design a vehicle location system based on the fusion of GNSS, INS, and LiDAR sensors. The system estimates the location through position, speed, and attitude together. The location-based on the LiDAR sensor shows the position and heading angle of the vehicle. Results show that the location incertitude with LiDAR decreases between \SI{5}{cm} and \SI{10}{cm} for both longitudinal and lateral location. Demir \textit{et al.}~\cite{Demir:2019} develop a framework for vehicle location that uses \SI{4}{x} LiDARs working simultaneously. Sensor readings are accumulated and merged into a scan accumulator module, which performs a normal distribution transform to analyze ambient variations using statistics on the point cloud distribution rather than point-to-point correspondence at accumulated data from each sensor. Results show that the maximum lateral and longitudinal error is \SI{10}{cm} and \SI{30}{cm}, respectively.

To mitigate adverse effects on vision-based sensors and laser-based sensors, some studies employ radar readings for localization applications. Abosekeen \textit{et al.}~\cite{Abosekeen:2019} estimate the vehicle location through a radar sensor for adaptive cruise control (ACC), in a navigation scheme that integrates GNSS/INS. The system performs raw radar measurement processing to determine the vehicle's estimated position and reduce the ground reflection effect and uses an EKF to combine the radar with a Reduced Inertial Sensor System (RISS). 

\smallskip\subsubsection{SLAM-based navigation}\label{subsubsec:SLAM-based} One concept of mapping used in robotic mobility is the well-known Simultaneous Localization and Mapping (SLAM). It represents a computational problem that tends to build or update a map of an unknown environment as soon as the vehicle moves, constantly monitoring the route~\cite{Martin:2014}. SLAM can use different exteroceptive sensors to which an algorithm is associated, depending on the scope and assumptions of the implementation. 

Vision-based approaches with stereo, monocular and thermal cameras are studied in~\cite{Lemaire:2007, Magnabosco:2013}. Basically, the authors combine stereo and monocular cameras with thermal cameras, to improve the detection of landmarks and to analyze the average relative error to the position of landmarks. However, visual location-based suffers from climatic changes, lighting, among others. Chiang \textit{et al.}~\cite{Chiang:2020} implement a navigation system using smartphone sensors. The system integrates GNSS/INS sensors and cameras. The authors implement the ORB-SLAM (Oriented FAST and Rotated BRIEF)~\cite{Rublee:2011} technique to process images. Data from the sensors that make up the system are merged through an EKF algorithm. The results show that the GNSS/INS system with integrated SLAM improves the accuracy of position and velocity, from 43\% to 51.3\%.

LiDAR-based approaches are not sensitive to ambient lighting, surface texture, as well as supporting long-range and wide field of view (FOV). Ghallabi \textit{et al.}~\cite{Ghallabi:2018} use lane markings to calculate vehicle location using multilayer LiDAR within a map. Line detection employs the Hough transform, as soon as a map-matching algorithm is implemented to validate landmarks within the location system. Javanmardi \textit{et al.}~\cite{Javanmardi:2019} propose a location system based on LiDAR multilayers and 2D vector map and planar surface map formats, which represent building, building footprints, and ground. The idea is to reduce the size of the map while maintaining location accuracy. A hybrid map-based SLAM system through a Rao-Blackwellized particle filter is proposed by Choi~\cite{Choi:2014}. Basically, LiDAR readings are filtered to classify and compare landmarks and to establish the location and mapping of the vehicle. Moras \textit{et al.}~\cite{Moras:2010} present a scheme to monitor moving objects around the vehicle and map the environment statically with LiDAR echo readings. The scheme implements a framework that merges a dual space representation (polar and cartesian), which defines the local occupancy grid and the accumulation grid. Accurate localization is a prerequisite for such a scheme.

Jose and Adams~\cite{Jose:2005} implement mmWave radar and formulate a SLAM problem to estimate target radar positioning and cross sections. Cornick \textit{et al.}~\cite{Cornick:2016} use a Localizing Ground Penetrating Radar (LGPR) that works through mapping and logging components. Ort \textit{et al.}~\cite{Ort:2020} use EKF to combine LGPR readings with wheel encoders and IMU sensor readings including magnetometer measurements.

\smallskip\subsubsection{Map tracking datasets}\label{subsubsec:map_tracking} To characterize maps in real-time, sensor readings are used to compose updated maps of the vehicle's surroundings. The mapping process occurs through object detection, where artificial intelligence techniques are implemented to identify various macro-areas, from lane markings to traffic signs recognition. As a result, the generation of maps in 2D and 3D is done through geometric and semantic layers. In addition to navigation, it is possible to analyze the dynamics and behavior of objects in the surroundings. Table\,\ref{tab:tracking_datasets} lists available datasets from experimental vehicles that collect data through exteroceptive sensors. All datasets include RGB cameras and LiDAR sensors. Rasterized maps in~\cite{Nuscenes:2019, Argoverse:2019, Lyft_Perception:2019} include roads, ground height and sidewalks, and vectorized maps include semantic layers like lane geometry, among others. Aerial map in~\cite{Lyft_Prediction:2020} is represented from the data encoded in the semantic map. Meanwhile,~\cite{Ford:2020} uses a ground plane and a 3D point cloud of non-roads data.

\begin{table}[!t]
\centering
\caption{Map tracking datasets available with exteroceptive sensors.}
\label{tab:tracking_datasets}
\begin{tabularx}{\columnwidth}{>{\centering\arraybackslash}p{1.8cm}>{\centering\arraybackslash}p{1.4cm}>{\centering\arraybackslash}p{0.6cm}>{\centering\arraybackslash}p{2.2cm}>{\centering\arraybackslash}p{0.6cm}}
\toprule
\textbf{Dataset}                                  & \textbf{Map type}       & \textbf{Layers}         & \textbf{Sensors}                        & \textbf{Size}       \\ \toprule
\multirow{2}{*}{Nuscenes~\cite{Nuscenes:2019}}             & \multirow{2}{*}{Raster}          & \multirow{2}{*}{11}                 &LiDAR/Radar /Camera/GPS/IMU              & \multirow{2}{*}{6\,h}       \\ \midrule
Argoverse~\cite{Argoverse:2019}  & Vector+Raster   & 2                 & LiDAR/Camera            & 290\,km       \\ \midrule
Lyft5~\cite{Lyft_Perception:2019}              & Raster          & 7                  & LiDAR/Camera                           & 2.5\,h     \\ \midrule
\multirow{2}{*}{Lytf5~\cite{Lyft_Prediction:2020}}   & HD\,Semantic & \multirow{2}{*}{7} & \multirow{2}{*}{LiDAR/Camera} & 1,118\,h    \\
                                           & Aerial      &                    &                               & 74\,km$^2$ \\ \midrule
\multirow{2}{*}{Ford~\cite{Ford:2020}}          & \multirow{2}{*}{3D}            & \multirow{2}{*}{2}        & LiDAR/Camera            & \multirow{2}{*}{66\,km} \\ 
&   &   &   /GPS/IMU &  \\ \bottomrule
\end{tabularx}
\end{table}

\section{Open Research Challenges}
\label{sec:challenges}

The utilization of exteroceptive sensors and their versatility in various telematics services and applications demonstrate their importance. Nonetheless, there are still open challenges, some of them inherent to the data acquired. In this section, we describe some of the areas which require further investigation. 

\smallskip\textit{Which data is important?}
Exteroceptive sensors onboard a Waymo vehicle generate up to \SI{19}{TB} of data per day~\cite{Waymo1,Tuxera_Waymo}. Clearly, determining which data is relevant becomes crucial for fast processing. Three strategies are useful to reduce the amount of data upstream:
\begin{enumerate}
    \item select and restrict the number of exteroceptive sensors;
    \item activate sensors only when necessary;
    \item degrade sensor precision (e.g., sampling frequency or image resolution).
\end{enumerate}

Some car manufacturers go further. Tesla renounces to use LiDAR sensors, claiming  they are unreliable~\cite{Forbes_Tesla}. As shown in Section~\ref{sec:services_and_applications}, different options and sensors are used to achieve the same function. On the other hand, sensor data fusion is  undeniably approach to achieve better precision and reliability. 

\smallskip\textit{Data processing.}
The analysis of sensor readings can have a high computational cost entailing response delay. 
In areas like vehicular safety and insurance telematics, where data analysis is used to detect critical events, response time is crucial.
With the commercialization of vehicles with different levels of autonomy, data analysis becomes more significant for the design of risk models.

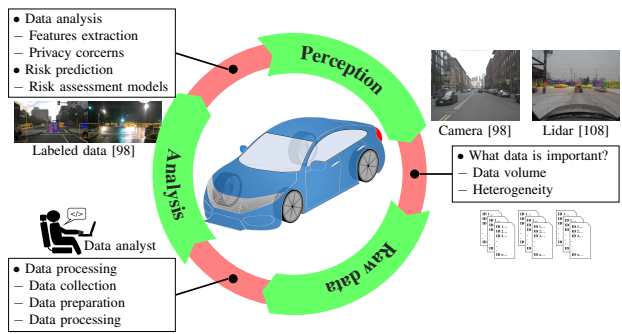
\begin{figure}[t]
\centering
\resizebox{0.47\textwidth}{!}{

\newcommand*{\mytextstyle}{\Large\bfseries\color{black!85}}
\newcommand{\arcarrow}[3]{%
   \pgfmathsetmacro{\rin}{2.5}
   \pgfmathsetmacro{\rmid}{3}
   \pgfmathsetmacro{\rout}{3.5}
   \pgfmathsetmacro{\astart}{#1}
   \pgfmathsetmacro{\aend}{#2}
   \pgfmathsetmacro{\atip}{5}
   \fill[color=green!60, very thick] (\astart+\atip:\rin) arc (\astart+\atip:\aend:\rin) -- (\aend-\atip:\rmid) -- (\aend:\rout) arc (\aend:\astart+\atip:\rout) -- (\astart:\rmid) -- cycle;
   \path[decoration = {text along path, text = {|\mytextstyle|#3}, text align = {align = center}, raise = -1.0ex},
      decorate
   ](\astart+\atip:\rmid) arc (\astart+\atip:\aend+\atip:\rmid);
}
\tikzstyle{data}=[align=left, rectangle, draw = black]
\begin{tikzpicture}
    \fill[even odd rule,red!50] circle (3.3) circle (2.7);
    \node at (0,0) [color = green, align = center]{{\includegraphics[width=0.25\textwidth]{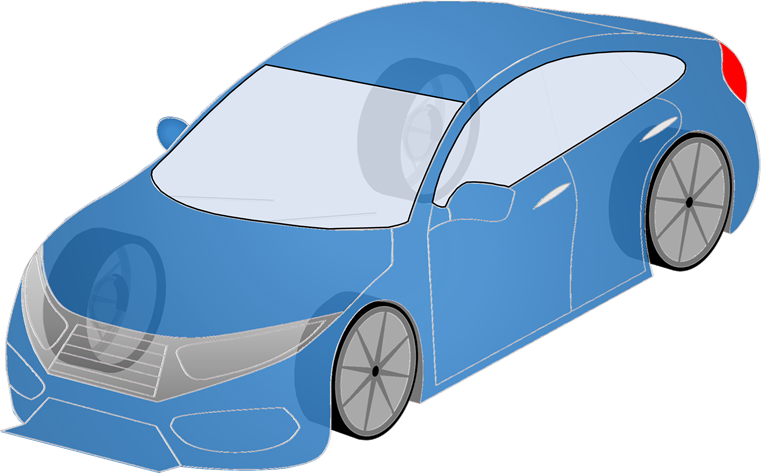}}};
    \node at (-5.6,-1.4) [color = green, align = center]{{\includegraphics[width=0.075\textwidth]{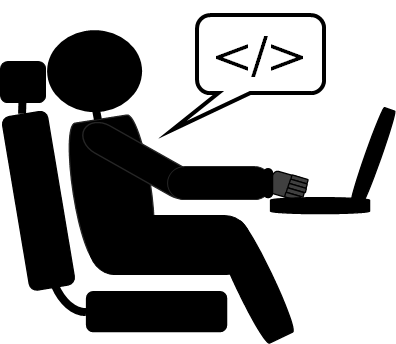}}};
    \node at (6,-1.5) [color = green, align = center]{{\includegraphics[width=0.15\textwidth]{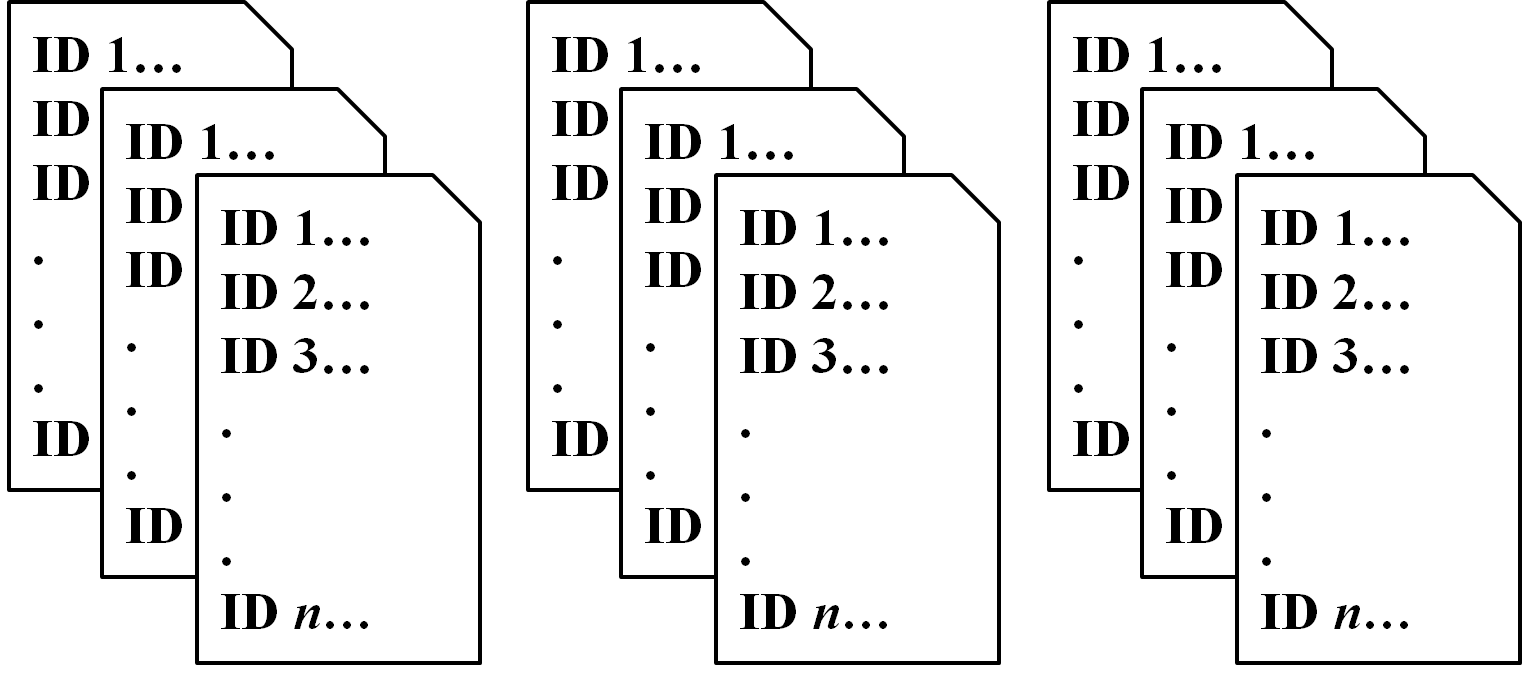}}};
    \node at (4.5,2.2) [color = green, align = center]{{\includegraphics[width=0.12\textwidth]{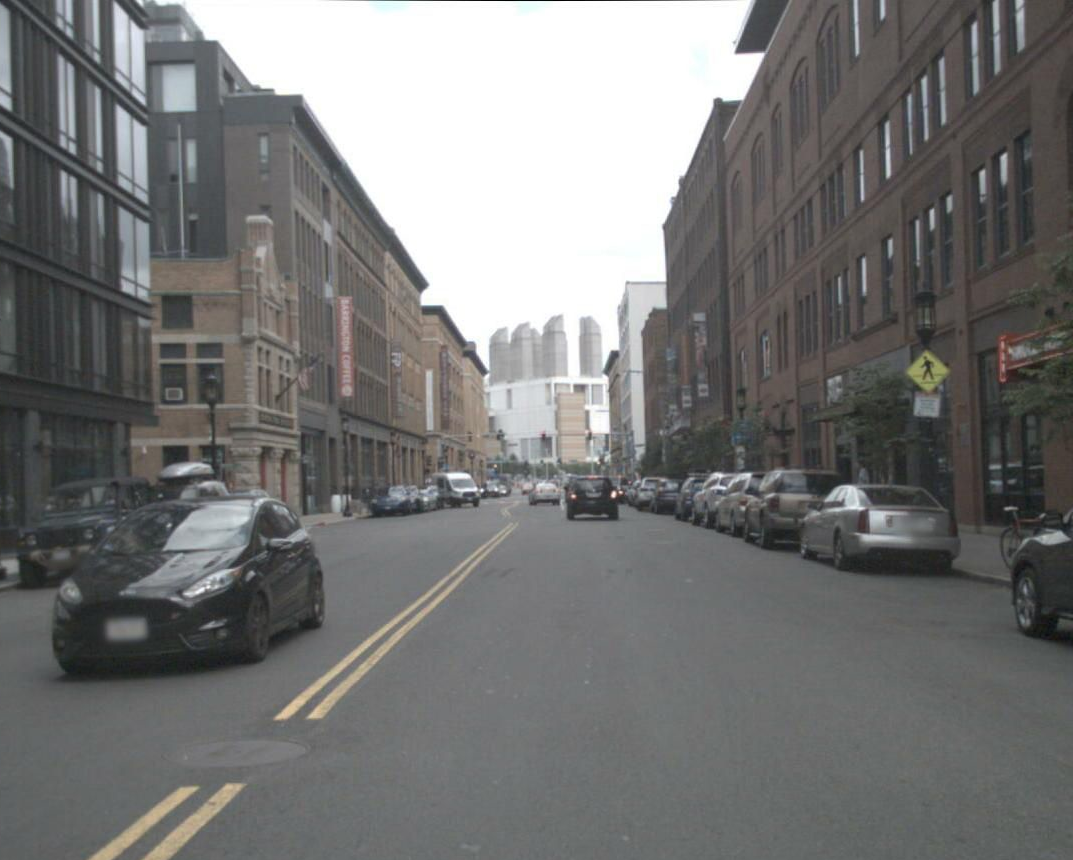}}};
    \node at (7,2.2) [color = green, align = center]{{\includegraphics[width=0.12\textwidth]{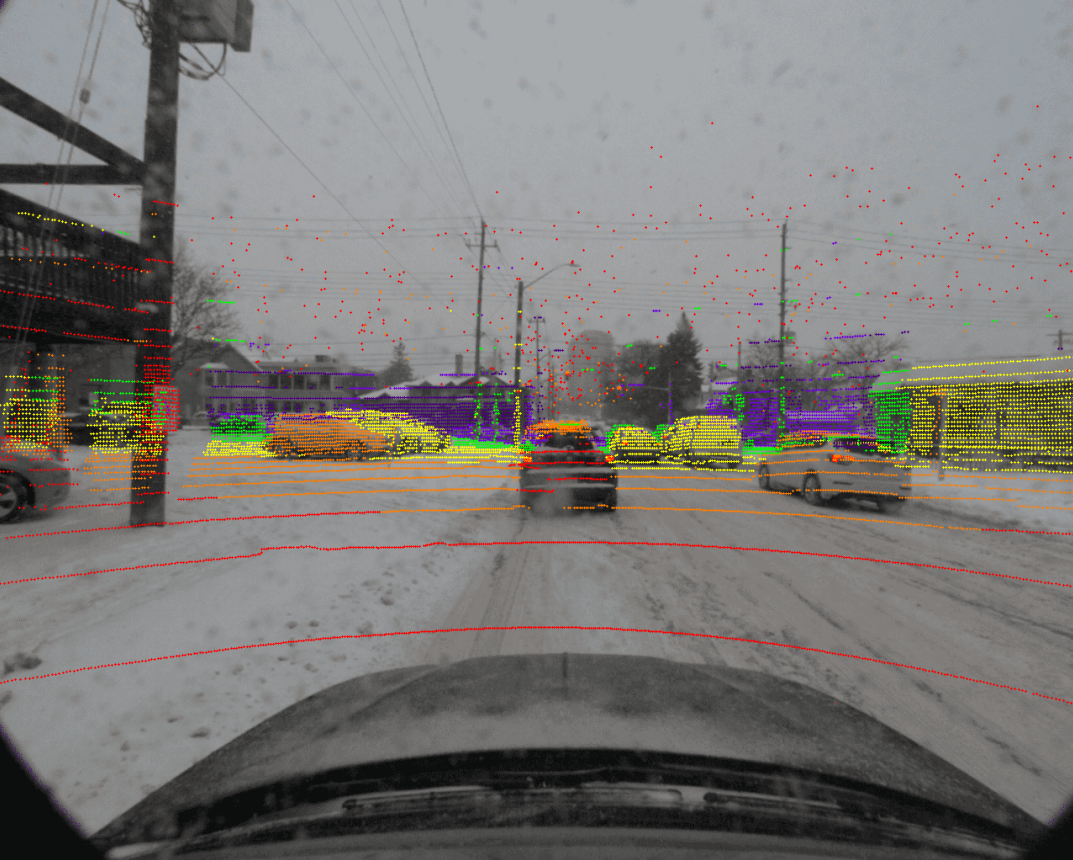}}};
    \node at (-5.1,1.3) [color = green, align = center]{{\includegraphics[width=0.20\textwidth]{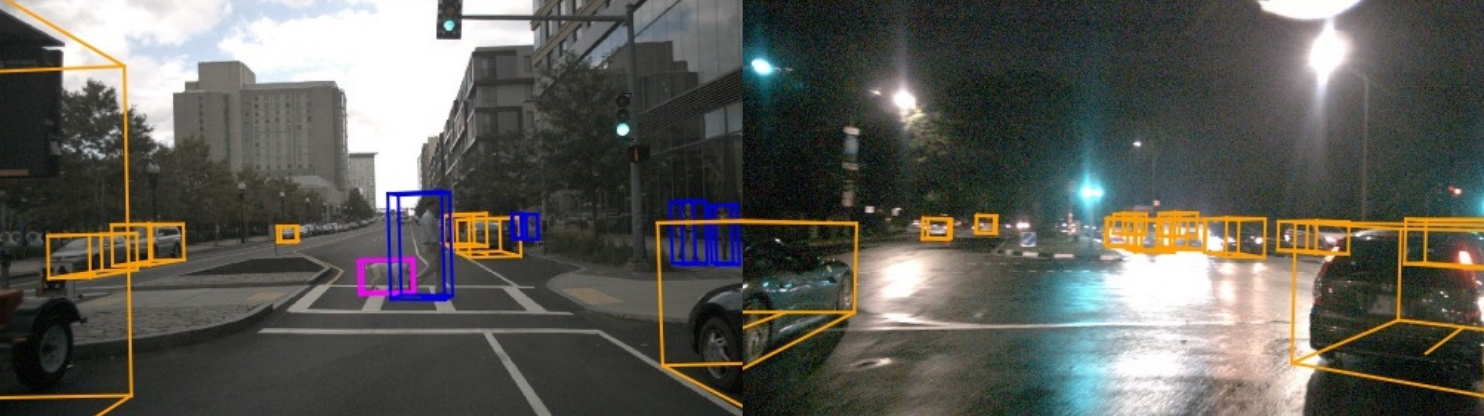}}};
    
    \arcarrow{100}{ 20}{Perception}
    \arcarrow{340}{260}{Raw data}
    \arcarrow{140}{220}{Analysis}
   
    \draw (6,0) node[data, text width=3.9cm] (b_perc){$\bullet$ What data is important? \\$-$ Data volume \\$-$ Heterogeneity};
    \draw (-5,3) node[data, text width=3.9cm] (b_anal){$\bullet$ Data analysis \\$-$ Features extraction \\$-$ Privacy corcerns \\$\bullet$ Risk prediction \\$-$ Risk assessment models};
    \draw (-5,-3) node[data, text width=3.9cm] (b_raw){$\bullet$ Data processing \\$-$ Data collection \\$-$ Data preparation \\$-$ Data processing};

    \filldraw[black] (3,0) circle (0.1cm) node[] (l_perc) {};
    \filldraw[black] (-1.5,2.598) circle (0.1cm) node[] (l_anal) {};
    \filldraw[black] (-1.5,-2.598) circle (0.1cm) node[] (l_raw) {};
    
    \foreach \f/\t in
        {b_perc/l_perc}
    \draw[black, very thick] (\f.west) -- (\t.east);
    
    \foreach \f/\t in
        {b_raw/l_raw, b_anal/l_anal}
    \draw[black, very thick] (\f.east) -- (\t.west);
    
    \draw (4.5,1.05) node[] {Camera~\cite{Nuscenes:2019}};
    \draw (7,1.05) node[] {Lidar~\cite{Canadian:2020}};
    \draw (-5.2,0.56) node[] {Labeled data~\cite{Nuscenes:2019}};
    \draw (-4.3,-1.8) node[] {Data analyst};
    
\end{tikzpicture}
}
\caption{Challenges involving vehicular telematics.}
\label{fig:challenges}
\end{figure}

This is compensated by Moore's law: hardware manufacturers are designing more and more performing computing systems which can be embedded in vehicles for data processing. Fig.~\ref{fig:challenges} shows different challenges involving data analysis.

\smallskip\textit{Security and privacy.}
A widely reported problem is data privacy and security~\cite{Tene:2012,Derikx:2016}. In the insurance market, for instance, the growth in the volume of telematics data and claims for coverage and compensation are essential to tailor services and insurance premium for each customer. On the other hand, such sensible data attracts cyber-attacks, forcing companies  to adopt extreme caution~\cite{Dambra:2020}. Currently, various works investigate methods to ensure and preserve the data integrity in insurance telematics~\cite{Pese:2017, Li_Privacy:2017, Zhou_Privacy:2019}.

An emerging technology that tackles privacy and security issues is blockchain. In a nutshell, a blockchain is a distributed database that stores indexes of transactions in a list of blocks that are chained to each other in a private and immutable manner~\cite{nakamoto2019bitcoin, Dorri:2017}. In addition to privacy, it may help preventing cyber frauds, as well as data manipulation by third parties.

\smallskip\textit{Risk assessment.} 
The evolution of autonomous vehicles creates a challenging scenario in term of risk assessment modeling for policymakers~\cite{SAE:2018}. The application of artificial intelligence in the data collected by the sensors raises a series of questions about the complexity of decisions, for example,  fairness and explainability. As a matter of fact, the transition to autonomous driving raises ethical and moral questions~\cite{Mordue:2020}. The insurance market requires delineating common points between responsibilities and ethics to establish policies associated with vehicle functionalities and legislation~\cite{Bellet:2019}.

\section{Conclusion}
\label{sec:conclusion}

This paper focuses on exteroceptive sensors, embedded or placed inside the vehicles, and their possible utilization for telematics services and applications like mobility safety, navigation, driving behavior analysis, and road monitoring. Such applications are of great interest both for the automotive and insurance industry as well as research, smart cities, drivers, passengers and pedestrians. Showing that exteroceptive sensors provide alternative and smart solutions when proprioceptive sensors are not available or just inconvenient, we provide to the reader a taxonomy of references for specific application areas and device types. Given the extensive literature, which grows with the development of autonomous vehicles, we have selected most relevant works based on their release date,  innovation, and feasibility. First, we have introduced the sensor classification and detailed specifications, advantages, and limitations of exteroceptive sensors considering their availability in OTS telematics devices. Moreover, we provide a report on existing available datasets for specific applications. Those are of paramount importance to the design of applications, especially on areas such as CNN for image processing, which demand large amounts of training data to perform well. We concluded the paper identifying open challenges and research directions: while sensors are becoming more precise and the sensor fusion more popular, the amount of data to process also increases at fast pace.

Other environmental sensing information can come from different channels such as communication with RSU sensors (V2I), or with other vehicles (V2V), or listening to specific streams on social networks. They are out of the scope of this paper though, as the works we refer to do not rely on a road infrastructure or on immeasurable data. Finally, the utilization of commodity devices in telematics grows steadily, smartphones \textit{in primis}, as they embed a large array of sensors.

\section*{Acknowledgment}
This study was financed in part by the Coordenação de Aperfeiçoamento de Pessoal de Nível Superior - Brasil (CAPES) - Finance Code 001, CNPq, FAPERJ, and FAPESP Grant 15/24494-8.

\bibliographystyle{IEEEtran}
\bibliography{exteroceptive.bib}

\ifCLASSOPTIONcaptionsoff
  \newpage
\fi
\vspace{-3em}
\begin{IEEEbiography}[{\includegraphics[width=1in,height=1.25in,clip,keepaspectratio]{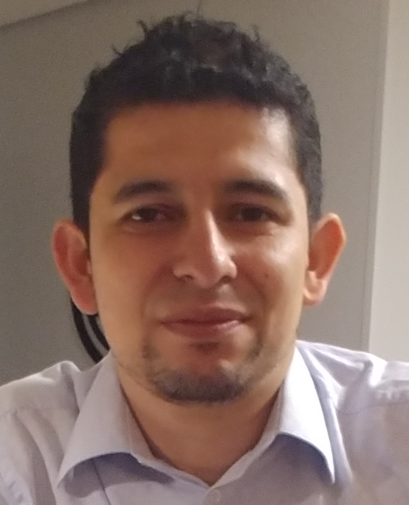}}]%
{Fernando Molano Ortiz} received bachelor degree from Universidad Católica de Colombia (UCC), Bogotá, in 2011, and his M.Sc. degree in Electrical Engineering from Universidade Federal do Rio de Janeiro (UFRJ), Brazil, in 2018. He is currently working toward Ph.D. degree at COPPE/UFRJ, in Brazil. His major research interests are on sensor networks, Internet of Things, smart mobility and vehicular networks.
\end{IEEEbiography}
\vspace{-4em}
\begin{IEEEbiography}[{\includegraphics[width=1in,height=1.25in,clip,keepaspectratio]{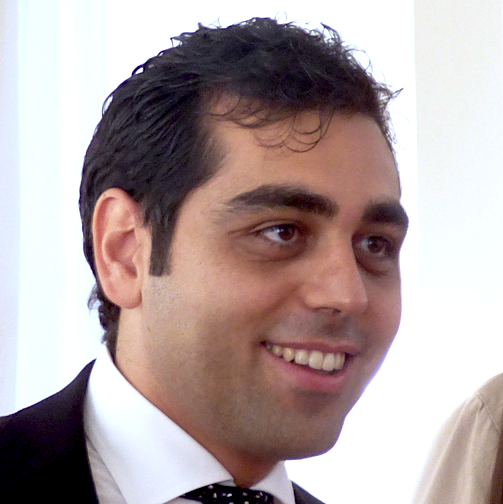}}]
{Matteo Sammarco} received his the BSc and MSc degrees in Computer Engineering from Università degli Studi di Napoli Federico II, Italy, in 2008 and 2010, respectively. In 2014 he received the Ph.D. in Computer Science from Sorbonne Université, France. He has worked as researcher at Télécom ParisTech and Laboratory of Information, Networking and Communication Sciences, both in France. Currently, he is research data scientist at Axa Group Operations, mainly working on machine learning applied to Safety, Smart Mobility and IoT.
\end{IEEEbiography}
\vspace{-4em}
\begin{IEEEbiography}[{\includegraphics[width=1in,height=1.25in,clip,keepaspectratio]{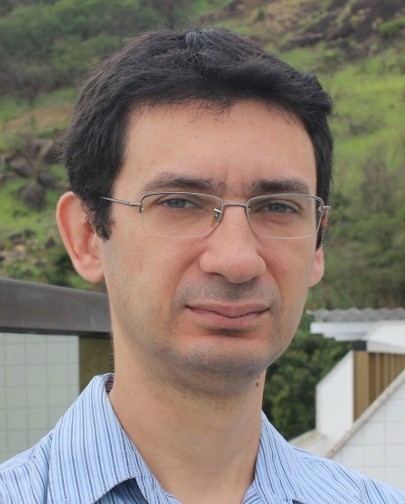}}]
{Luís Henrique M. K. Costa} 
received his Eng. and M.Sc. degrees in electrical engineering from Universidade Federal do Rio de Janeiro (UFRJ),  Brazil, and the Dr. degree from Universit{\'e} Pierre et Marie Curie (Paris 6), Paris, France, in 2001. Since August 2004, he has been an associate professor with Poli/COPPE/UFRJ. His major research interests are in the areas of routing, wireless and vehicular networks. 
\end{IEEEbiography}
\vspace{-4em}
\begin{IEEEbiography}[{\includegraphics[width=1in,height=1.25in,clip,keepaspectratio]{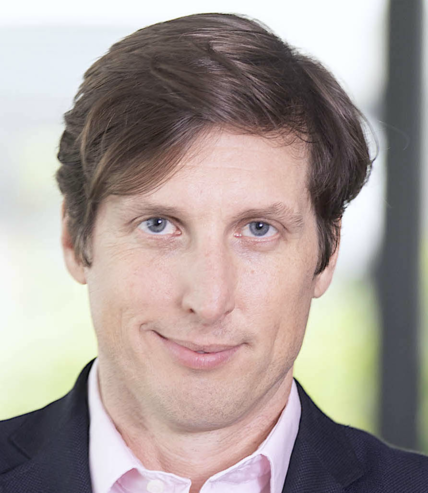}}]
{Marcin Detyniecki} studied mathematics, physics and computer science at the Sorbonne Université Paris. In 2000 he obtained his Ph.D. in Artificial Intelligence from the same university. Between 2001 and 2014, he was a research scientist of the French National Center for Scientific Research (CNRS). He has been researcher at the University of California at Berkeley and at Carnegie Mellon University (CMU) and visiting researcher at the University of Florence and at British Telecom Research labs.
He is currently Group Chief Data Scientist and Global Head of R\&D at AXA Group Operations. His work focuses on Machine Leaning, Artificial Intelligence, Computational Intelligence, Multimedia Retrieval, Fair and Transparent AI.
\end{IEEEbiography}




\end{document}